\RequirePackage{fix-cm}
\documentclass[smallextended]{svjour3}       
\smartqed  
\usepackage{graphicx}

\usepackage{tabularx,booktabs,ragged2e,float,adjustbox,makecell}

\usepackage[hyphens]{url}
\usepackage{hyperref}
\usepackage{textcomp}

\usepackage{balance}
\usepackage{soul}
\usepackage{color, colortbl}
\usepackage{tikz}
\usepackage{multirow}
\usepackage[english=american]{csquotes}
\usepackage{subfig}
\usepackage{tcolorbox}
\usepackage{graphics}
\usepackage{hyperref}
\usepackage{url}
\usepackage{amsmath,amssymb,amsfonts}
\usepackage{threeparttable}
\usepackage{adjustbox}
\usepackage{comment}

\usepackage{listings}
\definecolor{codegreen}{rgb}{0,0.6,0}
\definecolor{codepurple}{rgb}{0.58,0,0.82}
\usepackage[figuresleft]{rotating}
\usepackage{multirow}
\usepackage{listings}
\usepackage[english=american]{csquotes}
\usepackage{float}
\usepackage{soul}
\usepackage{adjustbox}
\usepackage{tcolorbox}
\usepackage{rotating}

\usepackage{caption}
\usepackage{subcaption}
\usepackage[numbers, sort]{natbib}
\usepackage[normalem]{ulem}

\newcommand{\Arghavan}[1]{\textcolor{orange}{{\it [Arghavan: #1]}}}

\begin{document}

\title{Bugs in Large Language Models Generated Code: An Empirical Study\thanks{This work was supported by: Fonds de Recherche du Québec (FRQ), the Canadian Institute for Advanced Research (CIFAR) as well as the DEEL project CRDPJ 537462-18 funded by the 
Natural Sciences and Engineering Research Council of Canada (NSERC) and the Consortium for Research and Innovation in Aerospace in Québec (CRIAQ), together with its industrial partners Thales Canada inc, Bell Textron Canada Limited, CAE inc and Bombardier inc.}
}

\titlerunning{Bugs in Large Language Models Generated Code}        

\author{Florian Tambon* \and
        Arghavan Moradi Dakhel* \and
        Amin Nikanjam \and 
        Foutse Khomh \and 
        Michel C. Desmarais \and 
        Giuliano Antoniol
}

\authorrunning{F. Tambon and A. Moradi Dakhel et al.} 

\institute{Florian Tambon
           \and
           Arghavan Moradi Dakhel
              \and
              Amin Nikanjam \and
            Foutse Khomh \and
            Michel C. Desmarais \and
            Giuliano Antoniol \at
            Polytechnique Montréal, Montréal, Canada\\
            \email{\{florian-2.tambon, arghavan.moradi-dakhel, amin.nikanjam, foutse.khomh, michel.desmarais, giuliano.antoniol\}@polymtl.ca}\\
            ``*'' authors contributed equally
}

\date{Received: date / Accepted: date}

\maketitle

\begin{abstract}
Large Language Models (LLMs) for code have gained significant attention recently. They can generate code in different programming languages based on provided prompts, fulfilling a long-lasting dream in Software Engineering (SE), i.e., automatic code generation. Similar to human-written code, LLM-generated code is prone to bugs, and these bugs have not yet been thoroughly examined by the community.
Given the increasing adoption of LLM-based code generation tools (e.g., GitHub Copilot) in SE activities, it is critical to understand the characteristics of bugs contained in code generated by LLMs. 
This paper examines a sample of 333 bugs collected from code generated using three leading LLMs (i.e., CodeGen, PanGu-Coder, and Codex) and identifies the following 10 distinctive bug patterns: \textit{Misinterpretations, Syntax Error, Silly Mistake, Prompt-biased code, Missing Corner Case, Wrong Input Type, Hallucinated Object, Wrong Attribute, Incomplete Generation, and Non-Prompted Consideration}. The bug patterns are presented in the form of a taxonomy.  
The identified bug patterns are validated using an online survey with 34 LLM practitioners and researchers. The surveyed participants generally asserted the significance and prevalence of the bug patterns. Researchers and practitioners can leverage these findings to develop 
effective quality assurance techniques for LLM-generated code. This study sheds light on the distinctive characteristics of LLM-generated code. 

\keywords{Large Language Models \and Bugs \and Software Testing \and Empirical Study}
\end{abstract}

\section{Introduction}
The recent surge in the development of Large Language Models (LLMs) tailored for code generation has garnered significant attention~\cite{wong2022exploring,imai2022github,sakib2023extending}. Transformer-based models such as Codex~\cite{chen2021evaluating} and Llama-2~\cite{touvron2023llama}, trained on large amounts of open-source code repositories, have demonstrated success in producing code across various programming languages, such as Python, Java, and C~\cite{yu2023codereval, Jin23, humanevalplus}. These models approach code generation as a transformation process, converting natural language descriptions (prompts) into executable programming language statements.

Positioned as potential AI pair programmers~\cite{MORADIDAKHEL2023111734,imai2022github,bird2022taking} in software projects, LLM-based code generators are poised to play a crucial role in the quality of the overall software project. However, similar to human-written code, LLM-generated code is prone to errors  \cite{MORADIDAKHEL2023111734}. Nonetheless, Asare et al. \cite{Asare-23} who recently examined vulnerabilities contained in LLM-generated code (using Copilot) compared to those in human-written code, reported that LLM models do not generate the same vulnerabilities as humans. In parallel, several studies \cite{li2024machines, kou2023model} reported that LLMs and human developers may not focus on the same part of the prompt to solve a coding task. This raises an important question: Do LLMs generate faults similar to those produced by human developers? This question is crucial because the effectiveness of popular quality assurance techniques like mutation testing depends on a precise characterization of faults occurring in the code under test \cite{Just-14}.
To illustrate the potential differences between bug patterns in LLM-generated code and those in human-written code, let's consider the example from Listing \ref{example1} which presents the code generated by Codex \cite{chen2021evaluating}, a well-known LLM to generate code, for the task of \enquote{returning the flags present in the Python ArgumentParser object} (shown in green as a docstring in the figure).

\begin{figure}[h!]
\begin{lstlisting}[language=python, caption={An example of a code generated by Codex \cite{chen2021evaluating}. The LLM is given the function signature and docstring (shown in green) as a prompt and should generate the code to address the task in the prompt. We labeled this buggy code as `` Non-Prompted Consideration (NPC)'' in our study.}, label={example1}]
def parser_flags(parser):
    '''
    Given an argparse.ArgumentParser instance, return its argument flags in a space-separated
    string.
    '''
    flags = []
    for action in parser._actions:
        if isinstance(action, argparse._SubParsersAction):
            for choice, subparser in action.choices.items():
                flags.append(choice)
                flags += parser_flags(subparser)
        else:
            flags.append(action.option_strings[0])
    return " ".join(sorted(flags))

\end{lstlisting}
\end{figure}

While the code seems reasonably correct at first glance, the attentive reader will notice in the last line (Line 14) that the LLM suddenly decided to sort the flags it extracted using the `sorted' function. This step, not requested in the prompt, leads to the code being faulty for the requested use case. This kind of snippet is interesting as it highlights two key points: 1) LLMs might generate a seemingly correct code at first glance, which could deceive non-experienced users particularly if no (or not enough) testing is performed, and 2) LLMs may add a non-prompted feature to the code, leading to an error, something that a human developer would rarely do, which shows that LLMs bugs might not be exactly similar to human-made bugs. 

While there exist studies that investigated bugs in LLM-generated code \cite{fan2023automated,liu23,pan2023understanding}, to the best of our knowledge, none of them examined the bug patterns in code generated by LLMs, when applied to real programming tasks (i.e., involving the implementation of functionalities). 

This paper fills this gap in the literature by examining the characteristics of bugs contained in code generated by three leading LLMs: CodeGen, PanGu-Coder, and Codex. The identified characteristics are then organized in a taxonomy and validated via a survey of practitioners and researchers. 
Our investigation is conducted using the following Research Questions (RQs):
\begin{itemize}
\item [\textbf{RQ1}] What are the characteristics of bugs occurring in code generated by LLMs for real-world project tasks?
\item [\textbf{RQ2}] To what extent are the identified bug patterns in LLM-generated code relevant for software practitioners and researchers working with LLMs?
\end{itemize}

We conducted our study using CoderEval \cite{yu2023codereval}, which is a dataset of 230 Python functions selected from 10 real-world open-source projects hosted on GitHub. Each of the selected LLMs was used to generate code for the different tasks described in CoderEval. We identified the bugs contained in the obtained code and manually analyzed the characteristics of a random sample of 333 bugs. This led us to organize our findings in a taxonomy composed of 10 categories of bugs: \textit{Misinterpretations, Syntax Error, Silly Mistake, Prompt-biased code, Missing Corner Case, Wrong Input Type, Hallucinated Object,
Wrong Attribute, Incomplete Generation, and Non-Prompted Consideration.} 
To assess the validity of the obtained bug patterns, we performed an online survey with 34 software practitioners and researchers working with LLMs. The obtained results show that respondents encountered the different bug patterns we observed, in a proportion similar to that obtained from our sample set. 



This paper makes the following contributions: 
\begin{itemize}
    \item We study bug patterns occurring in LLM-generated code and their prevalence.
    \item We propose a taxonomy from the characteristics of the bug patterns in LLM-generated code.
    \item We validate our findings using a survey and provide insights for researchers and practitioners.
    \item We make the dataset used in this study publicly available online \cite{rep-package} for other researchers and practitioners to replicate or build upon our work.
    \end{itemize}

\textbf{The rest of the paper is organized as follows.} We present the background of the study in Section~\ref{sec:background}. Next, we detail the methodology of the study in Section~\ref{methodology} for both the empirical study of bugs and the validation survey. We then report our empirical investigation of bugs in LLM-generated code in Section \ref{section:Taxonomy}. The validation survey results are presented in Section \ref{sec:validation}. We provide a discussion on our findings in Section \ref{sec:discussion}. Then, we discuss the related works and threats to the validity of this study in Section~\ref{sec:related_work} and Section~\ref{sec:validity}, respectively. Finally, we present the conclusion and future work in Section~\ref{sec:conclusion}.

\section{Background}\label{sec:background}
In this section, we introduce the three LLMs used in our study and describe the open coding methodology followed in this paper. 

\subsection{LLMs for code generation}


Different LLMs have been introduced for automatic code generation in Software Engineering (SE). One highly potent LLM is OpenAI's Codex~\cite{chen2021evaluating}, which has been extensively employed in diverse code-related SE tasks \cite{nguyen2022empirical,lemieux2023codamosa,chen2023teaching}. This closed-source GPT-based auto-regressive LLM~\cite{brown2020language} boasts up from 12 million to 12 billion parameters and undergoes fine-tuning on 54 million public repositories from GitHub. Notably, Codex is the model behind GitHub Copilot\footnote{{\url{https://copilot.github.com/}}}, an in-IDE developer coding assistant, proficient at generating code based on user-provided context with the maximum input length of 4,096 tokens. A recent addition to GitHub Copilot is ``Copilot Chat"\footnote{{\url{https://docs.github.com/en/copilot/github-copilot-chat}}}, which is tuned with human feedback for dialog use cases. Copilot Chat proves versatile for a broad range of code-related tasks. It excels at generating code fragments, describing code snippets in natural language, generating unit tests, and fixing buggy code, all tailored to the specific context of the task at hand.

PanGu-Coder \cite{christopoulou2022pangu} is an open-source pre-trained LLM for generating code from text.
This model is based on PanGu-$\alpha$ architecture \cite{zeng2021pangu}, a unidirectional decoder-only transformer with an additional layer for querying added on top of it. PanGu-Coder was trained in two stages using Python programs extracted from GitHub. 
In the first stage (unsupervised), the model is pre-trained on raw programming language data containing docstrings/inline comments written in natural language. Next, this model is fine-tuned to generate code from text in a second stage (supervised). 
Two versions of this model were published and leveraged for code generation recently \cite{zan2022neural}. These versions have respectively 317 million and 2.6 billion parameters, with a maximum input length of 1,024 tokens.

CodeGen \cite{nijkamp2022codegen} is a family of LLMs trained for text-to-code generation. Based on a decoder-only architecture, this family of models is designed for multi-turn program synthesis, in which a user engages with the LLM by gradually feeding specifications in natural language to obtain a corresponding program. 
This collaborative interaction allows the user, in conjunction with the LLM, to iteratively build and refine a program in multiple steps. CodeGen models were trained on a natural language corpus and code extracted from GitHub (i.e., a multilingual code dataset, and a dataset of Python programs). Various CodeGen models come with 350 million to 16.1 billion parameters and a maximum input length of 2,048 tokens \cite{nijkamp2022codegen}. CodeGen open-source models have been employed for code generation tasks in many recent studies~\cite{chen2022codet,zan2022neural}.


\subsection{Open Coding}
\label{sec:bk_OC}
Open coding, also known as postformed code, is a qualitative research method used to analyze textual data (e.g., survey responses, and interview transcripts). In open coding, researchers start with a relatively unstructured approach, inspecting the data line by line and generating codes or labels for themes and concepts identified in the data. The generated codes 
are subject to dynamic adjustments, i.e., they can be added, removed, or merged to develop a final codebook. This flexibility allows for a more adaptive and nuanced analysis compared to the rigid structure of preformed codes~\cite{seaman1999qualitative}. Open coding has proven to be more useful than the preformed code approach in SE studies~\cite{DL_faults,liu2023wants}. 

In open coding, two human reviewers independently start the process of coding on the same set of samples. During this phase, the reviewers explore patterns, similarities, and categories within the samples relevant to the quantitative variables~\cite{seaman1999qualitative}, i.e., characteristics or root causes of bugs in our study. They assign discrete categories to each sample. The two reviewers then engage in a discussion session to exchange and deliberate on their identified categories, forming the initial codebook.

In each discussion, a third reviewer, not involved in the coding process, intervenes in cases of diverging opinions between the two reviewers, to finalize the labels. The two reviewers continue to independently label the remaining samples and address similarly any conflicts that arise during subsequent discussion sessions~\cite{liu2023wants}. Through this iterative process, the two reviewers independently label all the samples. At the end of the labeling process, 
they report the agreement reached through these independent labeling processes before entering into any negotiations~\cite{liu2023wants}. Different studies in SE adapted this technique to conduct qualitative research for different purposes including constructing a taxonomy~\cite{DL_faults,liu2023wants}.

\section{Study Design}
\label{methodology}

\begin{figure*}
    \centering
    \includegraphics[width=\textwidth]{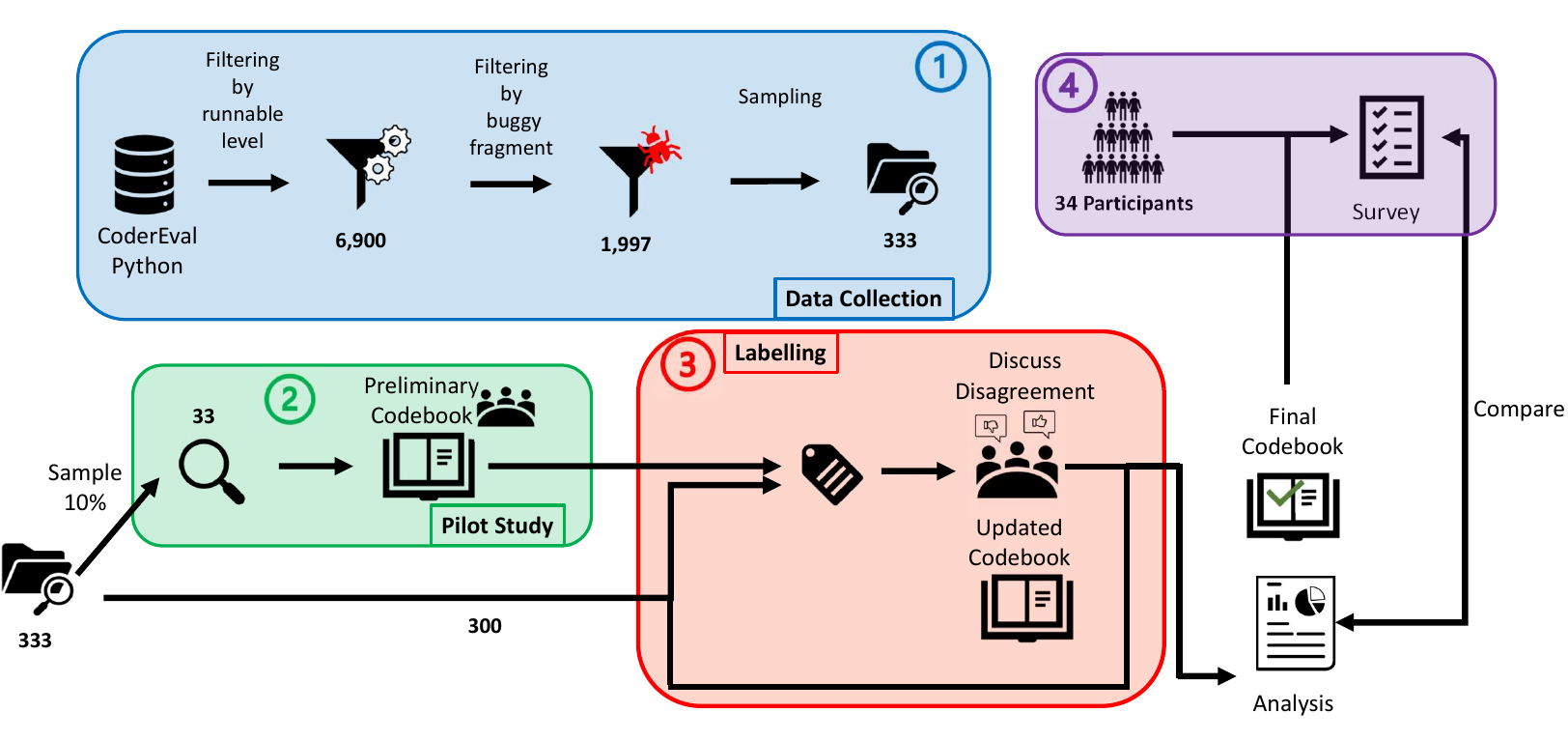}
    \caption{High-level view of the followed methodology.} 
    \label{fig:method}
\end{figure*}

In this section, we outline our methodology designed for investigating bugs in code generated by LLMs. We first describe our method for studying bug patterns in code generated by LLMs. Next, we present the methodology followed to validate the results of our empirical analysis. Specifically, we address the following research questions:
\begin{itemize}
    \item [\textbf{RQ1:}]What are the characteristics of bugs occurring in code generated by LLMs for real-world project tasks?
    \item [\textbf{RQ2:}]To what extent are the identified bug patterns in LLM-generated code relevant for software practitioners and researchers working with LLMs?
\end{itemize}
    
Figure \ref{fig:method} depicts the study methodology. We started by collecting data from a benchmark dataset (CoderEval) and selected a sample set. Next, we established a preliminary codebook by labeling 10\% of our sample set (pilot study). Subsequently, we iteratively completed the labeling process and our codebook. We finally organized our findings to obtain a taxonomy and surveyed practitioners to validate the bug patterns in our taxonomy. In the remainder of this section, we delve into further details of our methodology to answer each RQ.

\subsection{\textbf{RQ1: What are the characteristics
of bugs occurring in code generated by LLMs for real-world project tasks?}}

\subsubsection{Data collection}\label{sec:dt}
We used a benchmark dataset named CoderEval \cite{yu2023codereval} to conduct our study. This dataset consists of 230 Python functions from 43 open-source projects and 230 Java functions from 10 open-source projects selected from GitHub. In this study, we focused on Python functions as it is a very common and widely used programming language across different domains\footnote{\url{https://spectrum.ieee.org/top-programming-language-2020}}. Each function in CoderEval is constituted of the function signature, a docstring describing the function, and the oracle code (i.e., correct reference solution to the function). Additional information is available in this benchmark such as the context of the function (like other functions being used, and files/projects in which the function is contained). CoderEval also contains a human-written description of each function which was obtained by showing the oracle code to a group of developers without the docstring. Finally, for each of the functions, the CoderEval dataset contains 10 samples of code generated by each of the three LLMs under test for both types of prompt (docstring and human-written description). In our study, we focused on the code generated by LLMs using the docstring, which is a similar setting to other code generation benchmarks such as HumanEval \cite{chen2021codex} or MBPP \cite{austin21}.

At the time this study was started, CoderEval included code generated using three different LLMs\footnote{Note that at the start of this study, CoderEval contained the above-mentioned LLMs, which has since been updated later. As such, the version of the dataset we used in our study is: \url{https://github.com/CoderEval/CoderEval/tree/ec1177750cf10b5faa414a0e76d1430e75141a44}}: CodeGen~\cite{nijkamp2022codegen}, PanGu-Coder~\cite{christopoulou2022pangu}, and Codex~\cite{chen2021evaluating}. Two of these LLMs, CodeGen and PanGu-Coder, are open-source and have been used in previous studies that harness the power of LLMs for code generation tasks~\cite{chen2022codet,zan2022neural}. The third one, Codex~\cite{chen2021evaluating}, is the model behind Copilot~\footnote{\url{https://github.com/features/copilot}}. The code snippets generated by the LLMs were used to replace the oracle function sequentially. Upon replacement, the entry point code was executed, and the running output was captured, including return values, exceptions, errors, and console output. In case of no errors or exceptions, CoderEval compared the actual output with the expected output using unit tests. A difference between actual and expected output was considered a test failure. Conversely, an unexpected termination during execution was also treated as a test failure. The code snippets that led to a test failure were flagged as buggy in CoderEval. We used those flagged buggy samples in our empirical study.

The samples in CoderEval exhibit different ``runnable levels'', indicating the extent of dependency a function requires for effective execution. Out of the 6,900 Python samples in CoderEval (2,300 per LLM), our study was restricted to code fragments falling into one of the following runnable levels: \textit{``self\_runnable}'' (without external dependencies), ``\textit{slib\_runnable}'' (dependent only on Python's standard libraries such as ``os'' or ``json''), or ``\textit{plib\_runnable}'' (dependent only on Python's standard and public libraries such as ``numpy''). We made this decision both to reduce the number of potential samples as well as because higher ``runnable levels'' would be more likely to result in the generation of code that fails to address the intended task, as highlighted in CoderEval's paper \cite{yu2023codereval}. Indeed, the higher this level, the more information is needed to properly understand how the function should work. This information is less likely to be correctly encompassed in the docstring for higher ``runnable levels'', to the point that even a programmer not privy to the actual function would struggle to code the functionality. As such, any bugs in the code generated from the prompt of those categories would not necessarily reflect a weakness of LLM in generating code but just a lack of proper information. Therefore, we did not consider them to mitigate potential bias in our results. Finally, as our focus was on observing bugs generated by LLMs, we considered only those generated code fragments that were flagged as buggy by CoderEval. These filterings resulted in 1,997 buggy code samples.

To streamline the manual labeling effort, we opted to sample our data. At a 95\% confidence interval and an error rate of 5\%, this necessitated the analysis of 323 samples. We sampled \textbf{333} code fragments and ensured that the distribution of bugs across the three LLMs was balanced (highlighted as \raisebox{.5pt}{\textcolor{blue}{\textcircled{\raisebox{-.9pt}{1}}}} in Figure \ref{fig:method}).

\subsubsection{Manual labeling and classification}
As no previous categorization exists for bugs in LLM-generated code, we conducted a manual analysis of our buggy samples employing an open coding procedure~\cite{seaman1999qualitative}, similar to methodologies applied in prior studies on bug categorization~\cite{DL_faults,liu2023wants,Tambon2023silent}. The aim was to inductively construct bug categories in a bottom-up manner through manual analysis of buggy samples. This procedure allows us to identify various types of bugs based on our observations from the buggy samples. We used a shared document including the link to all buggy samples making sure that all reviewers could work together during the manual labeling process. The manual construction of the taxonomy involved three human reviewers, two of whom are senior Ph.D. students with experience in SE research. The third reviewer is a senior researcher with 10 years of research experience in the field of SE and AI. 

We developed the codebook of our bug patterns in multiple rounds. First, a preliminary pilot study was conducted. The first two authors independently coded 10\% of the sample set to reduce the subjective bias. Each coder was tasked to analyze each buggy sample and label it with one or more descriptive labels defined on their own, based on the bug pattern(s) observed in the sample. At the end of this round, the entire team convened to discuss the labels created, resolve conflicts, and merge the individual labels to create a consolidated set of categories, which served as the initial codebook for the next phase. In case of disagreement between the two coders, the third author would act as a tie-breaker (highlighted as \raisebox{.5pt}{\textcolor{green}{\textcircled{\raisebox{-.9pt}{2}}}} in Figure \ref{fig:method}).

In the second phase, the remaining buggy samples were divided into six parts, each containing 15\% of the remaining buggy codes, and we repeated the previous process: the two coders manually investigated all samples in each part independently, and then the entire team met to cross-check the labels and address conflicts while resolving discrepancies. If a new category emerged (i.e., a bug could not be classified under the existing categories) or a previous label was refined, the entire team gathered to discuss and incorporate the change into the codebook. We also made sure to re-label the affected samples from the previous rounds, if any. If a sample contained multiple types of bugs, it was classified into all identified bug patterns (i.e., using multiple labels). We had no constraint on the number of labels that could be assigned to a particular buggy snippet. The entire process required approximately 108 person-hours and resulted in a taxonomy comprising 10 bug patterns (highlighted as \raisebox{.5pt}{\textcolor{red}{\textcircled{\raisebox{-.9pt}{3}}}} in Figure \ref{fig:method}). 

Given that we did not have any prior-defined labels and since we do not have a predefined number of labels for every buggy code snippet in our dataset, it is not possible to compute inter-rater agreement levels; which is typical for open coding. However, after finalizing categories, we realized that \textbf{78.2\%} of the bugs were labeled similarly during independent labeling by the two reviewers before solving disagreements. Any conflicts were then thoroughly discussed until a 100\% agreement was reached~\cite{liu2023wants}. We retained all the initial labels and comments on our shared document to facilitate any subsequent discussions. We provide in Table \ref{tab:procedure_steps} each step of our process. These labels and comments are shared in our replication package available at~\cite{rep-package}.

\begin{table}[]
    \centering
     \caption{Coding process: initial agreements and conflicts throughout the rounds.}
    \begin{tabular}{cccccc}
         \toprule
         Round & Analyzed Samples & Initial Agreements & Conflicts & New Categories  \\
         \toprule
         Pilot & 33 & 26 & 7 & 8 \\
         1 & 50 & 37 & 13 & 1 (Wrong Attribute) \\
         2 & 50 & 42 & 8 & 1 (NPC) \\
         3 & 50 & 34 & 16 & 0 \\
         4 & 50 & 37 & 13 & 0 \\
         5 & 50 & 43 & 7 & 0 \\
         6 & 50 & 39 & 11 & 0 \\
         \bottomrule
         Total & 333 & 258 & 75 & 10 \\
         \bottomrule
    \end{tabular}
      \label{tab:procedure_steps}
\end{table}

\subsection{\textbf{RQ2: To what extent are the identified bug patterns in LLM-generated code relevant for software practitioners and researchers working with LLMs?}}

To assess the relevance of the bug patterns identified in \textbf{RQ1}, we surveyed software practitioners and researchers working with LLMs (\raisebox{.5pt}{\textcolor{violet}{\textcircled{\raisebox{-.9pt}{4}}}} in Figure \ref{fig:method}).

\subsubsection{Participants’ selection}\label{sec:participants}
To identify participants for our survey, we collected the contacts (email addresses) of GitHub users who collaborated on repositories that contain code generated using various LLMs. We followed the methodology proposed by Yujia et al.~\cite{fu2023security} to identify GitHub repositories containing code generated using LLMs. 
Yujia, et al.~\cite{fu2023security} employed various search keywords to collect GitHub repositories containing code generated by GitHub Copilot. We leveraged their shared dataset and broadened our list of participants, by adopting their methodology to collect additional repositories containing code-generated using 
Codex, Llama, Llama-2, CodeLlama, and ChatGPT. This process allowed us to collect a total of 113 repositories. Next, we employed the PyDriller~\cite{spadini2018pydriller} library to extract the email addresses of the collaborators on each repository. In total, we obtained 200 unique email addresses of practitioners who collaborated on GitHub repositories containing LLM-generated code. 

We complemented this list of practitioners with a list of researchers who have published on code generation using LLMs. To obtain this list of researchers, we searched relevant LLM-based code generation papers over Google Scholar\footnote{\url{https://scholar.google.com}} using two keywords: ``LLMs'' and ``code generation''. We collected the first 75 papers returned by Google Scholar and manually inspected each of them to ensure their relevance 
i.e., that they leveraged LLMs for automatic code generation. This inspection process allowed us to identify 56 relevant papers that focused on code generation using LLMs. Next, we extracted the email addresses of the authors of each of these papers (who are assumed to be researchers working on generating code with LLMs). We obtained a total of 182 emails after removing duplicates. Overall, we successfully sent the survey questionnaire to 382 unique email addresses. We discuss the details about the participants and the response rate in Section~\ref{sec:validation}.

We also posted our survey questionnaire 
on two relevant Reddit channels: \textit{LocalLLaMA} and \textit{MachineLearning}. Figure \ref{fig:reddit_post} in Appendix 
A shows our post on Reddit.

The questionnaire was dispatched along with a detailed message outlining the purpose, scope, and estimated completion time of the survey (which is around 10 minutes). We made it clear to the participants that the survey was anonymous, though participants had the option to provide their emails for additional communication and to receive a summary of the study.

\subsubsection{Survey with Practionners}\label{sec:val_meth}

The survey was open for two weeks, during which we sent a reminder one week after the initial email. Both the survey form and the anonymized responses that we collected are available in our replication package \cite{rep-package}. The survey is divided into two parts and is inspired by similar empirical studies of bug patterns~\cite{nikanjam2021DRL,DL_faults,Tambon2023silent}. In the first part, we asked demographic questions to participants. We then asked more specific questions about bugs in LLMs generated code regarding their frequencies and complexities associated with detecting or fixing the bugs (see Figure \ref{fig:survey} in Appendix B.1 for more details).


In the second part, we wanted to gather the participants' feedback on several aspects of each pattern of bug identified. To do so, we defined
the questions similarly for each pattern of bugs in the taxonomy. Specifically, for each pattern of bug, we present participants with a description of the bugs belonging to the pattern and provide an example code snippet; we use the same examples that will be presented in Section \ref{section:Taxonomy}. Next, we presented the participants with several questions to be answered using a 5-level Likert scale. The full questionnaire is presented in Appendix B.1 in Figure \ref{fig:survey_example}.

The questions start by asking about the frequency at which the participants encountered the bug pattern (ranging from 1-Never to 5-Always). Next, we defined the following concepts to the participants: 
\begin{itemize}
    \item[i.] Diagnosing: How hard would it be for the respondents to diagnose such an error in an LLM-generated code? Could they distinguish it easily or would it require running the code or using a debugger rendering the diagnostic hard? (1-Easy to 5-Hard) 
    \item[ii.] Complexity: Is the type of error rather trivial (i.e. not something a human developer with decent programming knowledge would do) or, on the contrary, does it denote some degree of complexity? (1-Trivial to 5-Complex)
    \item[iii.] Fixing: If the respondents had to fix such a mistake, would it be hard? Would it just require adjusting some part of the code (i.e. easy), or would it require extensive refactoring (i.e. hard)? (1-Easy to 5-Hard)
\end{itemize}
Participants were then asked to answer from their own perception/experience of each bug pattern.

The survey also contained optional comment fields at the end of each section and the end of the survey. Participants had the opportunity to provide suggestions such as additional information related to the bug patterns in the study or bug patterns not described in the study. 

\subsubsection{Analysis of Survey Results}\label{sec:analysis}

Once the results were collected for each bug pattern by the respondents, we aggregated the results (see Section \ref{sec:validation}). To do so, we followed a methodology used in similar studies \cite{Tambon2023silent} leveraging weighted average. We processed the results for the frequency/diagnosing/complexity/fixing of each bug pattern reported by the participants by multiplying the value of the Likert Scale with each percentage. That is to say, for instance, we would obtain a score of 2.79 for a bug pattern where respondents replied the following: 1 - (17.6\%), 2 (26.5\%), 3 (23.5\%), 4 (23.5\%), and 5 - (8.8\%). We also complemented the explanation of the results with comments from respondents relevant to the description of the results. The goal was to capture the perceived difficulty in fixing/diagnosing and the complexity of the different bug patterns according to practitioners. Additionally, we performed an analysis comparing the frequency of reported bug patterns among respondents with those in our sample set. We wanted to assess whether the bug patterns within our sample set (extracted from CoderEval) matched LLM practitioners' experience when dealing with bugs. The idea is to provide further evidence that our labeled categories are encountered in similar proportions by the respondents who used LLMs, not on a dataset such as CoderEval but as a tool in their development activities. Specifically, 
we evaluated if there was a correlation
 between the weighted average of the frequency of bug patterns reported by the participants and the proportion of each bug pattern found in the studied sample. To do so, we calculated the Spearman Rho between those scores and the distribution of bug patterns. The results are reported in Table \ref{tab:bug_taxonomy} (Section \ref{section:Taxonomy}).

 \section{Empirical Results}
In this section, we delve into the bug patterns obtained from our analysis as well as the respondents' answers from our survey. We examine the nature, frequency, and distribution of these bugs across the three different LLMs. All collected artifacts and data generated during our study are accessible in our replication package~\cite{rep-package}.

\subsection{\textbf{RQ1: What are the characteristics of bugs occurring in code generated by LLMs for real-world project tasks?}}\label{section:Taxonomy}

\subsubsection{Taxonomy}
\begin{figure*}
    \centering
    \includegraphics[width=\textwidth]{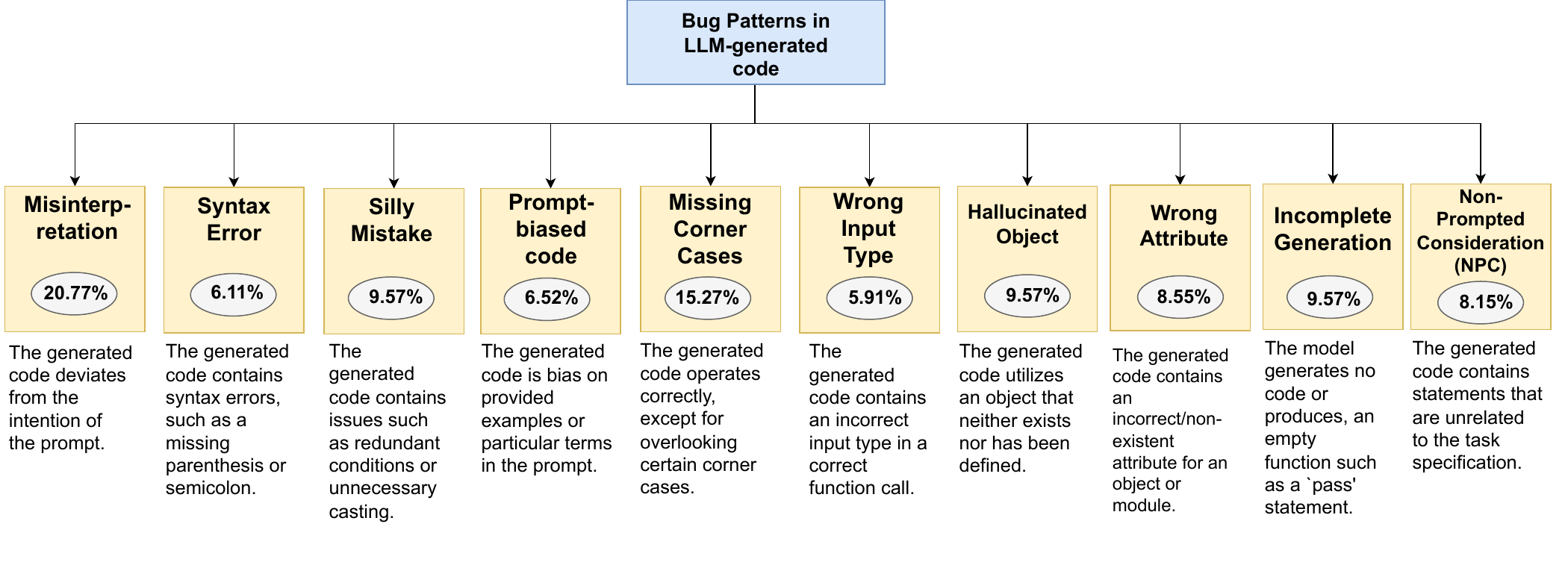}
    \caption{Final taxonomy of bug patterns in code generated by LLM. The number following each category represents the percentage of code samples assigned to that category during manual labeling.} 
    \label{fig:finaltaxo}
\end{figure*}

We organized our observations as a taxonomy. The taxonomy is organized into 10 categories. The full taxonomy is shown in Figure \ref{fig:finaltaxo} with the percentage of code samples assigned to each category during manual labeling.

In the following, we explain each of the 10 bug patterns observed; providing representative examples. 

\textbf{1. Misinterpretation:} Misinterpretation occurs when the generated code deviates from the intention of the prompt. This happens when the model struggles to capture the specifications of the task expressed in the prompt. In Listing \ref{snippet1}, you can find an example task, ``int\_to\_string'', along with its corresponding reference solution (top). The function declaration and the docstring at the beginning of the solution (lines 1 to 6) represent the prompt and the rest (lines 7 to 15) is the reference solution. A code generated by PanGu-Coder for this prompt is presented at the bottom of Listing~\ref{snippet1} (lines 17 and 18). This code is misaligned with the prompt since it simply treats the variable ``number'' as an index and returns the item at that index in the list ``alphabet''. However, the prompt describes a task to convert the variable ``number'' to a string using the given list ``alphabet''.  Consequently, we label the contained bug as \textit{Misinterpretation}. This bug pattern is among the most common in our analyzed dataset across all three models, with $20.77\%$ of buggy code samples categorized as \textit{Misinterpretation}.

\begin{figure}[h]
\begin{lstlisting}[language=python, caption={Reference Solution for the task ``int\_to\_string', and code generated by PanGu-Coder labeled as ``Misinterpretation''.}, label={snippet1}]
def int_to_string(number: int, alphabet: List[str], padding: Optional[int] = None) -> str:
    """
    Convert a number to a string, using the given alphabet.

    The output has the most significant digit first.
    """
    output = ``''
    alpha_len = len(alphabet)
    while number:
        number, digit = divmod(number, alpha_len)
        output += alphabet[digit]
    if padding:
        remainder = max(padding - len(output), 0)
        output = output + alphabet[0] * remainder
    return output[::-1]
_______________________________________________________________________________________________________________
def int_to_string(number: int, alphabet: str) -> str:
    return alphabet[number]
\end{lstlisting}
\end{figure}

\textbf{2. Syntax Error:} Any syntax error, such as a missing parenthesis or semicolon, is categorized as a \textit{Syntax Error}. Listing \ref{snippet2} showcases a code example with a syntax error in the first line generated by CodeGen for the same task, ``int\_to\_string'', as in Listing \ref{snippet1}. In this case, the LLM failed to complete the function declaration (adding the closing parenthesis) at the end of the function signature which resulted in a syntax error. It is worth noticing that \textit{Syntax Error} is one of the labels for the code example in Listing \ref{snippet2}. Although this bug pattern is not very common across the models in our study, one of the models in our analysis generated more buggy code with specific syntax errors, as further discussed in Section~\ref{sec:taxonomy_results}. 


\begin{figure}[h]
\begin{lstlisting}[language=python, caption={Code generated by CodeGen and labeled as ``Syntax Error''.}, label={snippet2}]
def int_to_string(number:
	"""
	Convert a number to a string, using the given alphabet.

The output has the most significant digit first.
	"""
	alphabet_string = ""
	for i in range(10):
		alphabet_string += chr(number % 10 + ord('A'))
		number //= 10

	return alphabet_string
\end{lstlisting}
\end{figure}

\textbf{3. Silly Mistake:} We labeled issues such as redundant conditions or unnecessary casting as \textit{Silly Mistake}. While these types of issues may not introduce a bug on their own, they have the potential to easily lead to one. In Listing \ref{snippet3}, there is an example generated by CodeGen where both the \textit{if} and \textit{else} conditions have the same statement. This is labeled as \textit{Silly Mistake}. This bug pattern ranks among the top three most common patterns across the models. 

\begin{figure}[h]
\begin{lstlisting}[language=python, caption={Code generated by CodeGen and labeled as ``Silly Mistake''.}, label={snippet3}]
if not logfile.exists():
	logging.basicConfig(
			filename=str(logf),
			format='%(asctime)s %(levelname)s %(name)s %(message)s',
			datefmt='%Y-%m-%d %H:%M:%S',
			level=logging.DEBUG,
		)
else:
	logging.basicConfig(
			filename=str(logf),
			format='%(asctime)s %(levelname)s %(name)s %(message)s',
			datefmt='%Y-%m-%d %H:%M:%S',
			level=logging.DEBUG,
		)
\end{lstlisting}
\end{figure}

\textbf{4. Prompt-biased code:} This issue occurs when the LLM excessively relies on provided examples or particular terms in the prompt while implementing a function and it sometimes hinders the generalization or correctness of the generated code. Listing \ref{snippet4} illustrates this issue. Lines 1 to 9 constitute the prompt, while lines 10 to 13 show the code generated by Codex for the task described in the prompt. The code should return the polygon for given vertices. The length of the vertices can be varied. However, lines 5 to 8 contain an example of vertices with length 4 and the list of its expected outputs, given to clarify the task provided in the prompt. In the code in Listing~\ref{snippet4} generated by Codex, the code is accurate only for the specific example provided in the prompt, when the length of the vertex is 4, and will fail in other cases. This code is labeled as a ``Prompt-biased code''. This type of pattern is less common across the models; however, CodeGen and Codex generated more buggy code biased on the prompt compared to PanGu-Coder. We discussed more details in Section~\ref{sec:taxonomy_results}.

\begin{figure}[h]
\begin{lstlisting}[language=python, caption={Code generated by Codex and labeled as ``Prompt-biased Code''.}, label={snippet4}]
def vertex3tuple(vertices):
    """returns 3 points for each vertex of the polygon. This will include the vertex and 
   the 2 points on both sides of the vertex:

        polygon with vertices ABCD
        Will return
        DAB, ABC, BCD, CDA -> returns 3tuples
        #A    B    C    D  -> of vertices
    """
    return [(vertices[-1], vertices[0], vertices[1]),
			(vertices[0], vertices[1], vertices[2]),
			(vertices[1], vertices[2], vertices[3]),
			(vertices[2], vertices[3], vertices[0])]

\end{lstlisting}
\end{figure}

\textbf{5. Missing Corner Case:} This bug occurs when the generated code operates correctly, except for overlooking certain corner cases. Listing \ref{snippet5} provides an example of this bug pattern along with its corresponding reference solution (top). Lines 17 to 24 represent the prompt and line 25 shows the code generated by Codex. The task involves implementing a function that checks if a given host is the localhost. Although the Codex-generated code checks for various representations of localhost, the reference solution is expected to verify additional potential options, such as \textit{socket.gethostname()}, which were present in the oracle given for the task described in the prompt. Listing \ref{snippet5} is labeled as \textit{Missing Corner Case}. This bug pattern is the second most common pattern with $15.27\%$ of buggy code samples categorized as \textit{Missing Corner Case}.

\begin{figure}[t]
\begin{lstlisting}[language=python, caption={Reference Solution for the task ``is\_local'' and code generated by Codex, labeled as ``Missing Corner Case''.}, label={snippet5}]
import socket
def is_local(host):
    """
    Checks if the host is the localhost

    :param host: The hostname or ip
    :return: True if the host is the localhost
    """
    return host in ["127.0.0.1",
                    "localhost",
                    socket.gethostname(),
                    # just in case socket.gethostname() does not work  we also try the following:
                    platform.node(),
                    socket.gethostbyaddr(socket.gethostname())[0]
                    ]
--------------------------------------------------
import socket
def is_local(host):
    """
    Checks if the host is the localhost

    :param host: The hostname or ip
    :return: True if the host is the localhost
    """
    return host in ['localhost', '127.0.0.1', '::1']
    
\end{lstlisting}
\end{figure}

\textbf{6.Wrong Input Type:} We use this label when LLM uses an incorrect input type in a correct function call. For example in Listing \ref{snippet6}, the task describes implementing a function to reduce a given list of classes to its ordered minimum equivalent. Lines 5 to 8 represent the code generated by CodeGen to address this task. In line 8, the Python built-in function \textit{min} is called on a List of ClassType inputs which raises an error. We labeled this buggy code example as \textit{Wrong Input Type}. It is worth noticing that this is not the only label for this buggy code example in Listing \ref{snippet6}. \textit{Wrong Input Type} is the least frequent bug pattern across the buggy code generated by three models.

\begin{figure}[h]
\begin{lstlisting}[language=python, caption={Code generated by CodeGen and labeled as ``Wrong Input Type''.}, label={snippet6}]
def minimalBases(classes):
	'''
	Reduce a list of base classes to its ordered minimum equivalent
	'''
	if len(classes) == 1:
		return classes[0]
	else:
		return min(classes)

\end{lstlisting}
\end{figure}

\textbf{7. Hallucinated Object:} We apply this label when the LLM hallucinates by attempting to utilize an object that neither exists nor has been defined. For instance, Listing \ref{snippet7} represents the code generated by Codex for the task ``make\_find\_paths''that involves transforming all the paths given in the find-paths variable into glob patterns. The generated function calls a function named ``find\_path\_to\_glob'' (line 14), which is not defined. \textit{Hallucinated Object} is among the third common bug patterns across the models. 

\begin{figure}[h]
\begin{lstlisting}[language=python, caption={Code generated by Codex and labeled as ``Hallucinatied Object''.}, label={snippet7}]
def make_find_paths(find_paths):
    '''
    Given a sequence of path fragments or patterns as passed to `--find`, transform all 
    path fragments into glob patterns. Pass through existing patterns untouched.

    For example, given find_paths of:

      ['foo.txt', 'pp:root/somedir']

    ... transform that into:

      ['sh:**/*foo.txt*/**', 'pp:root/somedir']
    '''
    return [find_path_to_glob(x) for x in find_paths]

\end{lstlisting}
\end{figure}

\textbf{8. Wrong Attribute:} This label is applied when the LLM utilizes an incorrect/non-existent attribute for an object or module. For example, in Listing \ref{snippet8}, there is code generated by PanGu-Coder for the same task presented in Listing~\ref{example1} in the introduction, named ``parser\_flag''. The prompt describes the task as ``Given an argparse.ArgumentParser instance, returns its argument flags in a space-separated string.'' The code generated by PanGu-Coder tries to use `flags' as an attribute of the input `parser', which is not a correct attribute according to the information provided in the prompt. The code sample in Listing \ref{snippet8} is labeled as ``Wrong Attribute''. This bug pattern is ranked as the fourth most common pattern, closely following the bug patterns in the third position.

\begin{figure}[h!]
\begin{lstlisting}[language=python, caption={Code generated by PanGu-Coder and labeled as ``Wrong Attribute''.}, label={snippet8}]
def parser_flags(parser):
    '''
    Given an argparse.ArgumentParser instance, return its argument flags in a 
    space-separated string.
    '''
    return''.join('--{0}'.format(f) for f in parser._flags)

\end{lstlisting}
\end{figure}

\textbf{9. Incomplete Generation:} This label is applied when the LLM generates no code or produces, for example, a `pass' statement or an empty function. This bug pattern also includes cases where the LLM stops completing the code in the middle of the function either because of token limits or because it reached an $<$end\_of\_sequence$>$ token without properly finishing the code. Listing \ref{snippet9} generated by CodeGen is an example of this category. \textit{Incomplete Generation} with \textit{Hallucinated Object} and \textit{Silly Mistake} are the third most frequent bug patterns we observed.

\begin{figure}[h!]
\begin{lstlisting}[language=python, caption={Code generated by CodeGen and labeled as ``Incomplete Generation''.}, label={snippet9}]
def write_configuration(config_filename, rendered_config, mode=0o600, overwrite=False):
	"""
	Given a target config filename and rendered config YAML, write it out to file. Create any
containing directories as needed. But if the file already exists and overwrite is False,
abort before writing anything.
	"""
	if os.path.exists(config_filename) and overwrite:
		print("Overwriting " +
		#

\end{lstlisting}
\end{figure}

\textbf{10. Non-Prompted Consideration (NPC):} This label is applied when the LLM adds statements to the code that are unrelated to the task specified in the prompt, leading to errors. For instance, in Listing \ref{example1}, which was presented as an example in the introduction, there is a code generated by Codex, for the task named ``parser\_flag''. The prompt involves ``Given an argparse.ArgumentParser instance, returns its argument flags in a space-separated string.'' As we already discussed in the introduction,  although the prompt does not specify sorting the flags, at line 14, the generated function sorts the list of flags before joining them, which is not in line with the given prompt. \textit{NPC} with a slight variation from the\textit{Wrong Attribute} type, stands at the fifth position with $8.15\%$ of buggy code examples across all models. 

\subsubsection{Analysis of the Bug Patterns} \label{sec:taxonomy_results}
The bug patterns have been defined based on the bug patterns observed in our studied samples. As mentioned in Section~\ref{methodology}, some of the buggy code samples are labeled in multiple categories due to the presence of different types of bugs or because a bug could be labeled as an overlap of multiple categories: \textbf{62.16\%} of the samples are labeled in a single category, \textbf{29.13\%} with two, \textbf{8.41\%} with three categories, and only one instance of buggy code is assigned four labels.

Table~\ref{tab:bug_taxonomy} presents the distribution of buggy code across different categories and by various LLMs. Some categories are more prevalent in specific models. For instance, in the case of a robust model like Codex, the most common bug pattern is \textit{Missing Corner Case} (\textbf{23.53\%} of the buggy samples). \textit{Missing Corner Case} occurs when the model generates code that addresses the task description but fails on one or a few exceptional test inputs due to a minor oversight, such as forgetting some particular cases. Listing \ref{snippet5} presents an example of this type. Conversely, in the case of relatively weaker and open-source models like PanGu-Coder or CodeGen, the most common category is \textit{Misinterpretation}. This bug pattern occurs when the code generated by the model does not align with the prompt and deviates from the intended purpose of the task.

\begin{table}[t]
  \caption{The distribution of bug patterns in the generated code by three different LLMs. We put in bold the top categories per LLM.} 
\centering
\begin{tabular}{l|ccc|c}
\specialrule{.1em}{.05em}{.05em}
\textbf{Bug Pattern} & \textbf{CodeGen} & \textbf{PanGu-Coder} & \textbf{Codex} & \textbf{Total} \\
\hline
 Misinterpretation & \textbf{22.84\%} & \textbf{24.11\%} & \textbf{15.03\%} & \textbf{20.77\%} \\
\hline
Syntax Error & \textbf{9.14\%} & 4.96\% & 3.27\% & 6.11\% \\
\hline
Silly Mistake & \textbf{14.72\%} & 4.26\% & 7.84\% & \textbf{9.57\%} \\
\hline
Prompt-biased code & 8.12\% & 3.55\% & 7.19\% & 6.52\% \\
\hline
Missing Corner Case & 5.58\% & \textbf{19.86\%} & \textbf{23.53\%} & \textbf{15.27\%}\\
\hline
Wrong Input Type & 6.60\% & 8.51\% & 2.61\% & 5.91\% \\
\hline
Hallucinated Object & 5.08\% & \textbf{14.89\%} & 10.46\% & \textbf{9.57\%} \\
\hline
Wrong Attribute & 7.61\% & 9.93\% & 8.50\% & 8.55\% \\
\hline
Incomplete Generation & \textbf{13.71\%} & 3.55\% & 9.80\% & \textbf{9.57\%} \\
\hline
NPC & 6.60\% & 6.38\% & \textbf{11.76\%} & 8.15\% \\
\specialrule{.1em}{.05em}{.05em}
\textbf{Total} & 100\% & 100\% & 100\%& 100\% \\

\specialrule{.1em}{.05em}{.05em}
\end{tabular}
\label{tab:bug_taxonomy}
\end{table}

Among all models, CodeGen has the highest number of buggy samples in the \textit{Silly Mistake} and \textit{Syntax Error} categories. Codex, as a stronger model, exhibits the highest number of samples in the \textit{Non-Prompted Consideration} category. This category pertains to situations where the model considers additional conditions or statements that were not requested in the prompt, leading to bugs. For example, the model might convert the final output of a function into an integer, which was not required in the prompt, or check if the output is a string and then apply further functionality (see Listing \ref{example1}). 

The third most common bug pattern in code generated by PanGu-Coder is \textit{Hallucinated Object}. This occurs, for example, when the model erroneously hallucinates the existence of a function that does not exist (see Listing \ref{snippet7}). Our observations reveal that when the model struggles to address the prompt, it sometimes resorts to calling an auxiliary function that is not defined, assuming that this function accomplishes the primary objective of the task. Usually, the function's name is generated based on the words used in the prompt.
 
When considering all three studied models, \textit{Misinterpretation} is the most common bug pattern (\textbf{20.77\%}). The second place belongs to \textit{Missing Corner Cases}, \textit{Hallucinated Object}, with \textit{Incomplete Generation} and \textit{Silly Mistake} following closely in the third position. It is noteworthy that the last three mentioned bug patterns should rarely be observed in code written by humans. For example, human developers typically do not frequently call undefined functions (\textit{Hallucinated Object}) or access attributes on objects that do not exist (\textit{Wrong Attribute}). Some of these bug patterns can be detected by Python Linters embedded in various IDEs, such as VS Code, which further reduces the likelihood of encountering them in human-written code. For the \textit{Silly Mistake} category, a common mistake is when the model generates an if/else condition but uses the same statement in both branches, which is also a rare mistake in human-written code.

\begin{tcolorbox}[colback=blue!5,colframe=blue!40!black]
\textbf{Findings 1:} While buggy samples in CoderEval exhibit \textit{Misinterpretation} and \textit{Missing Corner Cases} bug patterns (that would lead to an assertion error during unit testing), there are also a non-trivial number of more severe bug patterns such as \textit{Hallucinated Object} or \textit{Wrong Attribute}. Those mistakes are not typical of human developers, either because they show strange behavior (e.g., \textit{Silly Mistake} with duplicated if/else condition) or because they would be easily detected using an IDE.
\end{tcolorbox}

\begin{figure}[t]
\centerline{\includegraphics[width=1\textwidth]{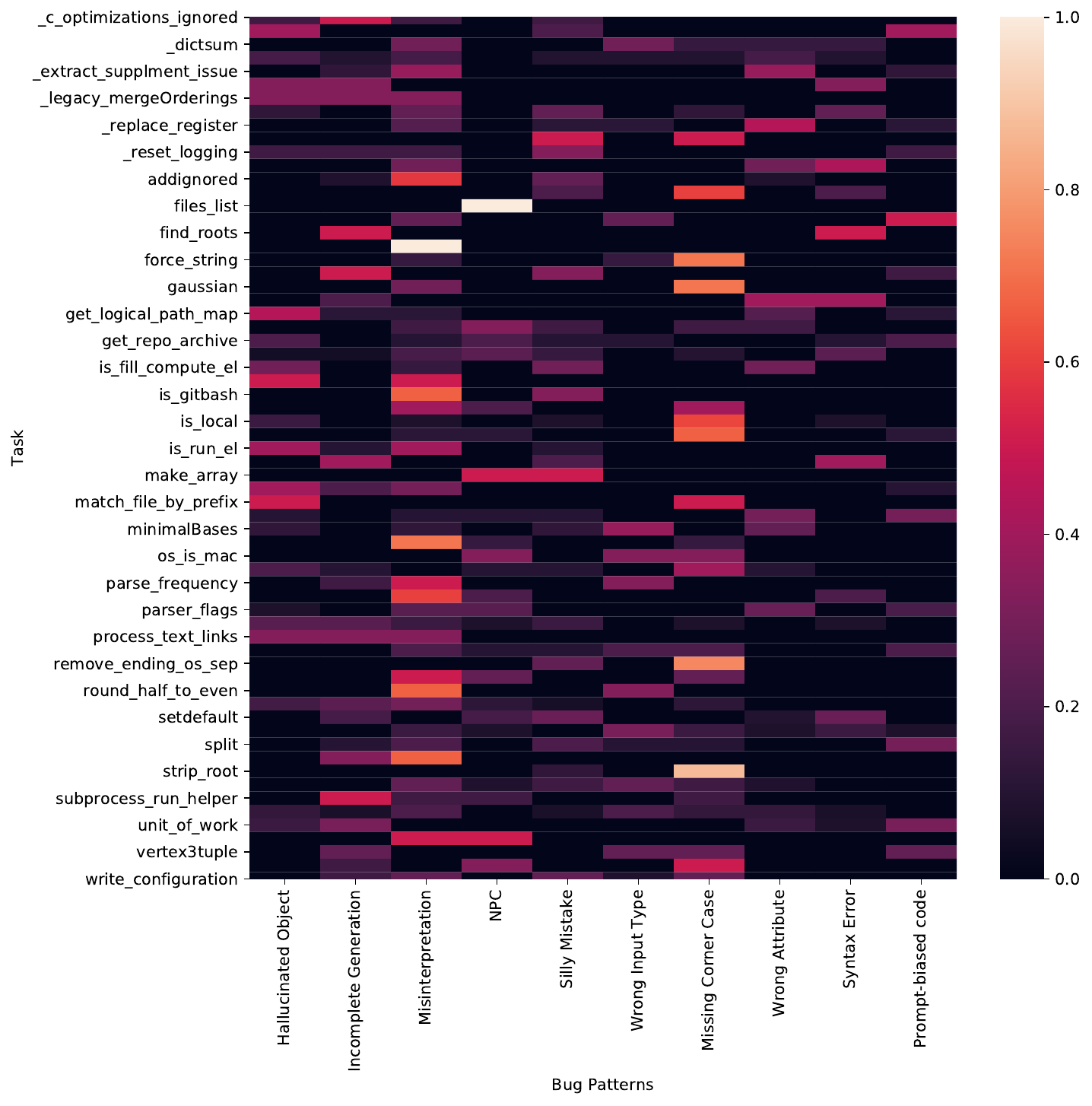}}
\caption{The heatmap illustrates the distribution of bug patterns on various tasks across all LLMs. The values are normalized over lines per task. The brighter areas represent a higher number of code samples of a specific task categorized into a particular bug pattern. The black areas indicate the lack of code samples in the bug pattern for a specific task.}
\label{fig:heatmap}
\vspace{-1em}
\end{figure}

To further study bug patterns in LLM-generated code on the CoderEval benchmark, we look into what types of bugs are present, for each task in the benchmark, across LLMs. The goal is to identify if some tasks in the dataset are more likely to lead to a specific bug pattern across all LLMs which could indicate some potential shortcomings in the prompts or if bug patterns are evenly distributed across different tasks in the CoderEval benchmark. To do so, we counted in our sample set, for each task, the number of times a particular bug pattern was made, regardless of the employed LLM. We then normalized values per task to know what bug patterns per task was more prevalent. Results are presented as a heatmap in Figure \ref{fig:heatmap}. 

Overall, the various bug patterns are evenly distributed across different tasks, as indicated by the darker areas, less than 0.5, in the heatmap, Figure~\ref{fig:heatmap}. This suggests that the tasks are not biased toward a particular bug pattern that we identified. However, the few very light spots on the heatmap (greater than 0.9) show the majority of code samples for a specific task, categorized under a particular bug pattern. Based on our results, only two tasks, all the buggy code snippets were classified under a single bug pattern. For example, for the task named ``fix\_namespace\_prefix\_w'' where all the buggy code snippets generated by different LLMs are categorized as \textit{Misinterpretation}. This classification stems from the fact that the prompt associated with this task, ``Convert text that defaults to `w:st=`' '  to `w-st=`' ', does not accurately represent the functionality revealed by the oracle and corresponding test cases. However, all models generated at least one correct solution based on the same prompt, which passed all the test cases. 


Additionally, as shown in Figure \ref{fig:heatmap}, only for a few tasks, the majority of bugs are concentrated in a specific type. For example, in the ``force\_string'' task, where up to 72\% of the bug patterns generated by different models are categorized as \textit{Missing Corner Case} due to an additional newline character in the prompt. The prompt for this task reads, ``This function returns the bytes object corresponding to `obj' / in case it is a string using UTF-8.''. The content after the backslash appears as a new line in the prompt, creating ambiguity in the prompt's objective for all LLMs. In the buggy code snippets, the function attempts to return the bytes object corresponding to `obj'. In cases where `obj' is a string, it decodes it to UTF-8. However, upon checking the oracle, it becomes evident that the intended purpose of the prompt is to return the corresponding UTF-8 if `obj' is an instance of a byte object.

As discussed in Section~\ref{sec:dt}, our result is limited to tasks with runnable levels no higher than \textit{``plib\_runnable''} (dependent only on Python's standard and public libraries, such as ``numpy''). The goal of this limitation is to mitigate potential biases in our results that may arise from the lack of information on dependencies not covered in the prompt. However, to further our analysis, we assess whether the three selected runnable levels in our sample set cause any bias towards specific bug patterns. To conduct this analysis, we report the distribution of bug patterns in different runnable levels of tasks across LLMs, as shown in Table~\ref{tab:bug-level}.

The results, as indicated in Table~\ref{tab:bug-level}, show a varied distribution of bug patterns across different runnable levels. Similar to the result in Table~\ref{tab:bug_taxonomy}, the \textit{Misinterpretation} category contains the maximum number of buggy code snippets in each dependency level, with the second-highest occurrences being \textit{Missing Corner Case} for \textit{``slib\_runnable''} and \textit{``plib\_runnable''}, and \textit{Wrong Attribute} for \textit{``self\_runnable''}. While higher runnable levels are more likely to cause generating buggy code by LLMs, as highlighted in CoderEval benchmark \cite{yu2023codereval}, the results in Table~\ref{tab:bug-level} do not demonstrate that increasing the dependency level of the task up to \textit{``plib\_runnable''} impacts the bug patterns in code snippets for most of the bug patterns.

We observed only the following trend in the case of three bug patterns: \textit{Missing Corner Case} (an increasing trend with runnable level) and \textit{Prompt-biased code}/\textit{Non-Prompted Consideration} (a decreasing trend with runnable level). For the \textit{Missing Corner Case}, the results suggest that increasing the runnable level might make the model more likely to miss some expected cases because the prompt might lack information on how the library is expected to be used to tackle the problem. On the contrary,  the amount of \textit{Prompt-biased Code} and \textit{NPC} increased when the runnable level decreased  (a decreasing trend with runnable level) since the models might rely more on the details in the prompt for tasks in the CoderEval dataset with a lower level of dependencies on the libraries. Similar trends are not observable for other types of bugs at different runnable levels. Therefore, considering the runnable levels in our dataset, our proposed bug patterns are not influenced by a specific runnable level.




\begin{table}[t]
  \centering
  \caption{Distribution of bug patterns across different runnable levels (levels of task dependencies). } 
  \label{tab:bug-level}
  \resizebox{\textwidth}{!}{
  \begin{tabular}{l|ccc|c}
    \specialrule{.1em}{.05em}{.05em}
    \textbf{Bug Pattern} & \textbf{self\_runnable} &  \textbf{slib\_runnable} & \textbf{plib\_runnable} &\textbf{Total}
    \\
    \hline
   Misinterpretation & \textbf{16.29\%} & \textbf{24.86\%} & \textbf{23.53\%} & \textbf{20.77\%} \\
    \hline
   Syntax Error & \textbf{7.69\%} & 4.32\% & 5.88\% & 6.11\% \\
    \hline
   Silly Mistake & 8.60\% & \textbf{11.35\%} & 8.24\% & 9.57\% \\
    \hline
   Prompt-biased Code & 8.14\% & 6.49\% & 2.35\% & 6.52\% \\
    \hline
    Missing Corner Case & 10.41\% & \textbf{17.84\%} & \textbf{22.35\%} & \textbf{15.27\%} \\
     \hline
  Wrong Input Type & \textbf{6.79\%} & 4.86\% & 5.88\% & 5.91\% \\
    \hline
  Hallucinated Object & 9.95\% & 9.73\% & 8.24\% & 9.57\% \\
     \hline
  Wrong Attribute & \textbf{13.12\%} & 2.16\% & \textbf{10.59\%} & 8.55\% \\
     \hline
   
    Incomplete Generation & 9.50\% & 10.27\% & 8.24\% & 9.57\% \\
    \hline
    
    NPC & 9.50\% & 8.11\% & 4.71\% & 8.15\% \\

    \specialrule{.1em}{.05em}{.05em}
    \textbf{Total} & 100\% & 100\% & 100\% & 100\% \\
    \specialrule{.1em}{.05em}{.05em}
  \end{tabular}
  }
  \vspace{-10pt}
\end{table}

\begin{tcolorbox}[colback=blue!5,colframe=blue!40!black]
\textbf{Findings 2:} The bug patterns are evenly distributed over 
different tasks in the CoderEval dataset.  The bug patterns are also evenly distributed over different runnable levels (dependency levels) of tasks in this dataset. However, our results reveal that the ambiguity present in the description of a task in this dataset may impact the distribution of buggy samples within a specific task between different bug patterns across different LLMs. 
\end{tcolorbox}

\subsection{\textbf{RQ2: To what extent are the identified bug patterns in LLM-generated code relevant for software practitioners and researchers working with LLMs?}}
\label{sec:validation}

The survey was open for two weeks resulting in 34 answers, i.e., we achieved a return rate of 8.9\%, which is in line with many SE surveys conducted outside specific companies \cite{Nardone23,nikanjam2021DRL}. We first give results of the questions concerning the demographics and experience of the participants. We then detail results regarding the frequency of reported bug patterns, as well as the analysis of the reported difficulty of diagnosing/fixing and the perceived complexity of the bug patterns (see Table \ref{table:table-survey-taxonomy}).

\subsubsection{Participants’ demographics and experience with LLM bugs}

All the participants answered the demographic questions. We obtain the following answers:

\begin{itemize}
    \item For the academic field: 12 Ph.D. students, 4 Researchers, 8 Undergraduate/Graduate students and 1 Lecturer. For the industry field: 6 Developers, 2 Data scientists and a CTO.
    \item ChatGPT was used by \textbf{31} participants. \textbf{67}\% of the participants reported only non-open-source LLMs such as ChatGPT, Bard~\cite{chowdhery2023palm} or Claude~\cite{claude2023}. \textbf{10} participants cited open-source LLMs such as LLama or CodeGen.
    \item In terms of programming language used for LLMs, besides Python, the participants responded: \textbf{47\%} for JavaScript, \textbf{27\%} for C++ and Java and \textbf{24\%} for C. These languages are among the most used programming languages according to different programming language indexes \cite{gitIndex, tiobeIndex}. Other programming languages were reported in smaller proportion: Rust (6\%), HTML/CSS (6\%), SQL (9\%), C\# (6\%), or Go (3\%). Three respondents (all PhD students) reported exclusively using LLMs only to program in Python language. 
\end{itemize}

Note that the sum of responses for LLMs and programming languages goes above the number of responses as one could provide multiple answers.

Before asking participants specific questions about the categories of bugs, we collected information about their general experience dealing with bugs in LLM-generated code. 
\textbf{80\%} of the participants reported that they often or always experience bugs when using LLMs to generate code. No respondent said they \textit{never} faced bugs in LLM-generated code. We asked the participants about the complexity of the bugs they encountered in their LLM code generation activities and \textbf{68\%} of them mentioned that the bugs had a medium difficulty to fix, with only \textbf{21\%} of participants reporting that their encountered bugs were  \textit{easy} to fix. 
We also asked the participants how they proceeded to fix the encountered bugs and 
\textbf{30\%} of them reported that they fixed the bugs manually by reading the stack traces and googling the error messages. The remaining participants reported using a combination of manual inspection with re-prompting the LLM to correct the mistake. No participant declared only using re-prompting based on the error message, similar to approaches presented in~\cite{schafer2023adaptive}. This highlights the fact that debugging LLM-generated code is still predominantly a manual task. We believe that the bug patterns proposed in this paper are an important first step toward the development of efficient automatic debugging tools for LLM-generated code. 

\subsubsection{Frequency of bugs patterns}

\begin{sidewaysfigure}
    \centering
    \includegraphics[width=\textwidth]{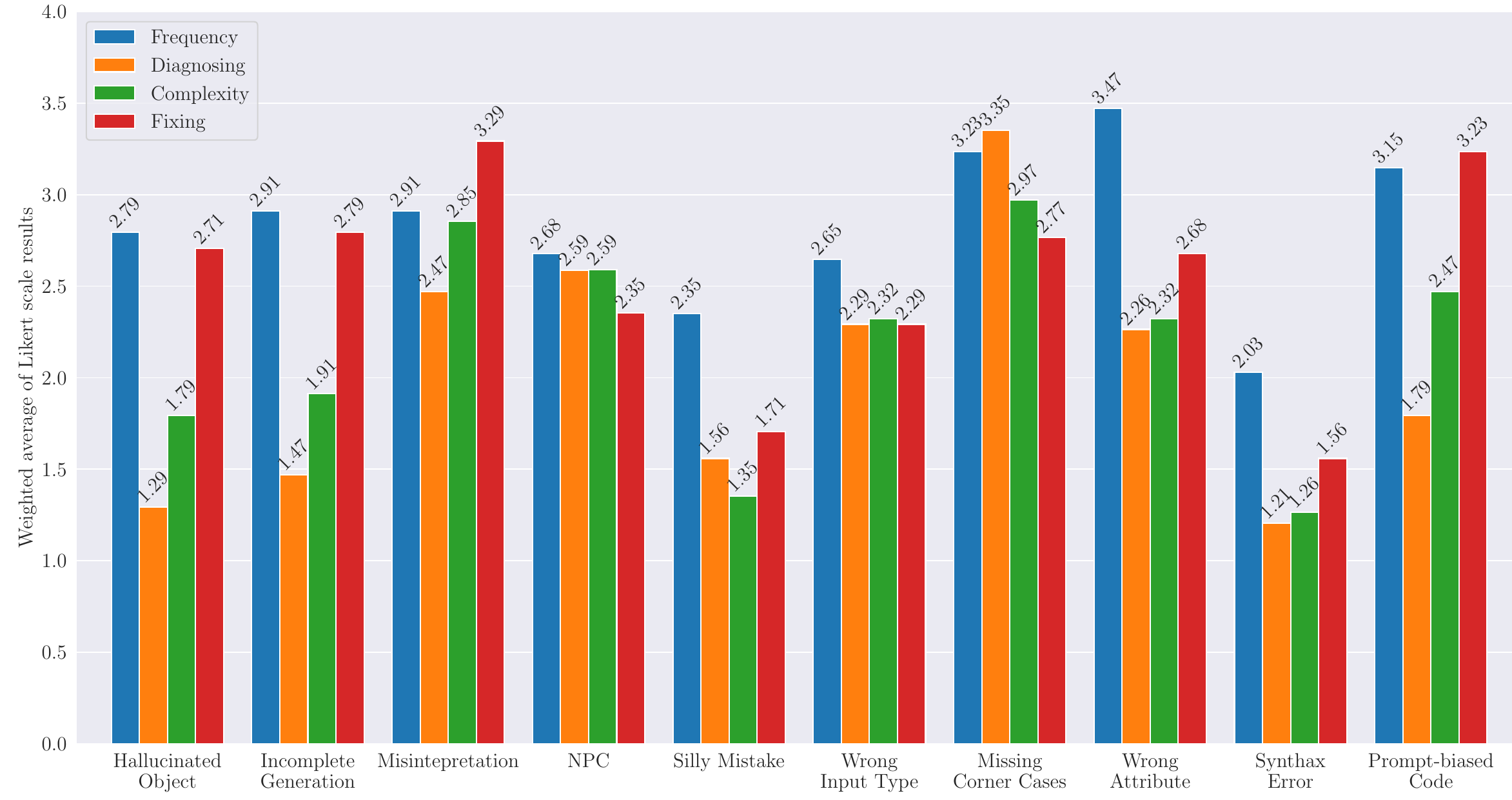}
    \caption{Aggregated results of the validation survey. Questions related to the frequency of encounter of bug patterns, the difficulty to diagnose and fix them as well as the complexity of the bug. We highlight in bold the highest number in each category for each bug pattern. 1 represents never/easy/trivial/low and 5 represents always/hard/complex/high.}
    \label{fig:bar_plot}
\end{sidewaysfigure}

Figure \ref{fig:bar_plot} summarizes the obtained aggregated results in the \enquote{Frequency} column. A table with the results for the answer per Likert scale score of the respondents is also available in the Appendix \ref{secA3}. For each category of bug, we report the following: the frequency at which participants declared they encountered the bug pattern, what is the perceived difficulty in diagnosing such a bug pattern, what is the perceived complexity of the bug pattern (i.e., is it a trivial mistake a competent developer would not do or does it denotes some complexity), and what is the perceived difficulty to fix such a bug pattern. For each of those metrics, we report the weighted average results giving the proportion of respondents that answered 1/2/3/4/5 given the Likert scale as described in Section \ref{sec:analysis}. For instance, the \textit{Missing Corner Cases} bug pattern has a score of \textbf{3.23} and is among the most frequent bug patterns encountered by participants. All categories in the taxonomy have been encountered and acknowledged by the participants. The most encountered categories are \textit{Wrong Attributes}, \textit{Prompt-biased Code}, or \textit{Missing Corner Cases} with an aggregated score above $3$. Thus, participants report that they often encounter bugs from these categories. One participant reported that they observed \textit{Prompt-biased Code} bugs when asking LLMs to generate code using a prompt with several examples of unit tests: \enquote{\textit{I face this problem when docstring given too many examples, and the prompt requires LLMs to pass at least all the test cases}}. The categories of bugs that were encountered the least are \textit{Syntax Error} and \textit{Silly Mistake} with scores respectively of \textbf{2.03} and \textbf{2.35}. Thus, most participants reported that they never or rarely encountered these two categories of bugs. 
This distribution is in line with the results obtained in RQ1, for Codex (one of the most popular LLMs) using the CoderEval dataset: \textit{Syntax Error}, \textit{Silly Mistake}, and \textit{Wrong Input Type} are the least frequent bug categories in Codex-generated code (see Table \ref{tab:bug_taxonomy}). Moreover, the \textit{Missing Corner Case} is the most frequent bug category in code generated using Codex. To confirm this observation, we conducted a correlation analysis as described in Section \ref{sec:val_meth} between the frequency of occurrence of bug patterns reported by respondents and the frequency of bug patterns observed in our sample set. The highest correlation is obtained for Codex (\textbf{$\rho$ = 0.47}, medium correlation) which seems plausible as we have seen at the beginning of this section that a large majority of the respondents worked using ChatGPT and/or Codex. The $rho$ for PanGu and CodeGen are \textbf{0.28} and \textbf{-0.18} respectively. As such, the frequency of the bug patterns reported by the respondents is similar to the distribution of the bug patterns sampled for the Codex model in our sample set. 

We do note some discrepancies such as \textit{Prompt-biased Code} or \textit{Wrong Attribute} type being more prevalent in respondents' answers than what we observed in CoderEval or \textit{Misinterpretation} not being as prevalent. This can be explained by the type of prompt used to generate code in CoderEval. Indeed, most of the prompts were extracted from real projects so information in the docstring might be limited which impacts the LLMs' generation (\enquote{\textit{Usually more outlining or instructions can be added after a docstring to help guide the model}}). This can be, for instance, because the developers who designed the code understood it. Thus, they put a limited docstring as it was sufficiently understandable for them, yet it might not be enough for an LLM. Similarly, CoderEval's prompt generally does not include an example of the task at hand which might limit \textit{Prompt-biased Code} prevalence. Nonetheless, even with a limited docstring, some errors such as \textit{Wrong Attribute}, \textit{Hallucinated Objects} or \textit{Wrong Input Type} are unexpected as they are more linked to expertise in a programming language and library used rather than an understanding of the task at hand.

\subsubsection{Diagnosing, Complexity, and Fixing of bug patterns} 

Figure \ref{fig:bar_plot} summarizes the obtained results in the \enquote{Diagnosing}, \enquote{Complexity} and \enquote{Fixing} columns which were defined to participants (see Section \ref{sec:val_meth}). A table with the results for the answer per Likert scale score of the respondents is also available in the Appendix \ref{secA3}. In terms of diagnosing the bugs, most bug patterns were reported by the respondents as easy to diagnose (score $< 2$), especially as several respondents pointed out that: [the bugs are] \enquote{\textit{Easy to diagnose when using an IDE with real-time code checking}}. According to our survey respondents, the only bug patterns that are not as easy to diagnose are \textit{Misinterpretation}, \textit{Non-Prompted Consideration}, and \textit{Missing Corner Cases}. In the case of \textit{Non-Prompted Consideration}, a respondent highlighted that: \enquote{\textit{I have observed some difficult cases that non-requested behaviour is in the middle of the generated code}}. \textit{Missing Corner Cases} is the only category reported as being quite hard to diagnose by respondents (score $ > 3$). This result could be expected since the \textit{Missing Corner Cases} category encompasses code that will compile correctly, but break unit tests, as highlighted by one of the participants: \enquote{\textit{It is a risk to become complacent to LLM code generation because sometimes code looks reasonable (in this case). Still, you might be unaware of all the ways it could fail}}. Another respondent emphasized the importance of efficient test cases when addressing bugs from the \textit{Missing Corner Cases} category, nothing that: \enquote{\textit{It is very dependent on the evaluation test case set. From its short natural language description, like docstring, it's hard to cover all its possible corner cases}}.

Regarding how likely a developer is to make such a mistake (i.e., \enquote{complexity} of the bug patterns), respondents mostly consider the identified bug patterns to have low complexity, with \textit{Syntax Error} and \textit{Silly Mistake} types being the least complex. This adds to our intuition that most bug patterns in LLM-generated code are different from human-made bugs. Indeed, as most developers use IDE, \textit{Syntax Error}s or similar mistakes are pretty easy to catch. Similarly \textit{Silly Mistake}s, such as the example presented in Listing \ref{snippet3}, are not a bug that developers would likely do, and, even if done, they are easy to detect. On the contrary, \textit{Misinterpretation}s and \textit{Missing Corner Cases} bug patterns are more complex and more similar to a type of error that a developer would make. This is highlighted by the respondents, as both categories have the highest aggregated score. One survey participant commented that: \enquote{\textit{For example, I would very much forget about IPv6 (don't we all?:)) and miss that corner case myself. So it is both my lack of knowledge and implicitly trusting the LLM output that can make this such an insidious bug to diagnose and solve}}. Thus, the complexity of such bugs could arise from both the nature of the error and developers' excessive reliance on LLMs, which amplifies the impact and intricacy of the issue.

Regarding the difficulty of fixing the bugs from the different categories, the survey participants 
consider the bugs to be moderately difficult to fix; 
mirroring their responses in the first part of the survey, when we asked about the difficulty of fixing LLM-generated bugs. Only \textit{Syntax Error}s and \textit{Silly Mistake}s are considered easy to fix with an aggregated score of $< 2$. Bugs from the \textit{Misinterpretation} category are reported to be more difficult to fix than the other categories by the survey's participants. This result is consistent with the observation we made during the labeling phase. \textit{Misinterpretation} bugs generally lead to a code that is way different compared to the prompt specification. In that case, re-prompting or extensive manual effort is needed to fix the generated code fragment. For some bug categories, like \textit{Hallucinated Objects}, participants suggested that providing additional information in the prompt can often help the LLMs to correct the mistake: \enquote{\textit{Most often time after providing a header or function definition, the model can correctly create the method or function}}. However, in several cases, especially for bugs in the \textit{Misinterpretation} category, it is not enough to provide additional information: \enquote{\textit{It depends on whether you know what you are expecting the code to look like. If so, this is easy to spot, and probably you're better off manually writing the code. I have tried to prompt-engineer, but I find that it's a waste of my time. Sometimes, you will get code generated that misinterprets the prompt, but maybe you are not aware of what the generated solution looks like so you have to read the generated code intently or figure out the library calls that are generated to understand that it misunderstood the prompt}}.

Finally, as we reported in Section \ref{section:Taxonomy}, some LLMs like Codex (and so Copilot), tend to generate additional code that is not explicitly requested in the task's description. This may be linked to the different settings of the LLM: an IDE plugin for Copilot/Codex or a web browser/standalone for CodeGen and PanGu-Coder. This was echoed by one of the respondents of our validation survey who also observed this kind of behavior being more prevalent in Copilot-based models compared to the others: \enquote{\textit{There is a difference in LLM-generated code quality between using an in-IDE plugin such as copilot, and using a browser-based tool such as chat gpt or google bard. I find that copilot has more of a tendency to recommend code that I don't need, or code that is redundant; [...]}}.

\begin{tcolorbox}[colback=blue!5,colframe=blue!40!black]
\textbf{Findings 3:} The surveyed practitioners encountered the different bug patterns in our empirical results. In particular, their reported frequency of bug patterns correlates with the frequency observed for \textit{Codex} in our sample set. They reported that bug patterns are generally easy to diagnose and fix but it requires manual effort. Those LLMs bug patterns are generally different from human-made bugs. \textit{Misinterpretation} and \textit{Missing Corner Case} bug patterns are the only exceptions: being harder to fix and diagnose. as well as denoting faults that are more akin to human-made. 

\end{tcolorbox}

\section{Lessons Learned}\label{sec:discussion}

This study shed light on the nature of bugs occurring in code generated using LLMs. We observed that 
bugs contained in LLMs-generated code differ from bugs in human-written code. 
Our analysis and the feedback received from survey participants also reveal that LLM-generated code can appear correct at first glance and deceive users depending on the complexity of the code and the experience of said users. To the light of our manual labeling and validation survey, in the following, we provide several lessons learned for both LLMs' users and researchers.

\subsection{Lessons Learned for LLM Users}\label{sec:practitioners}

\textbf{Limited information in the prompt can drastically lead the model astray:}

CoderEval's tasks are composed of a function signature with a docstring extracted from actual GitHub repositories. While this might be enough to some extent for a human, LLMs seem to struggle with this somewhat limited information. One issue we observed in our sample set concerns the library or even the Python version. Indeed, there is little control over which version of a given programming language or library the LLMs will use in the generated code if it is not indicated explicitly in the prompt. We came across examples of generated code that would only work for a certain version of Python while the prompt and the oracle code were pointing to another version. For instance, when parsing XML files, the LLMs would sometimes use the function \enquote{getchildren()} which is deprecated for certain versions of Python. This is something that some respondents in our survey reported: (\enquote{\textit{Sometimes, the LLM would generate functions that are deprecated. It is not necessarily hallucinating, it was just trained on an older version of the Python package you are using}}). Interestingly, by looking at other samples of the same task not necessarily in our sample set, we found that the usage of this function by LLMs was not systematic and some samples were correctly using the expected (more recent) function. Hence, having an older version in the generated code can not only be explained by deprecated training data. Similarly, on tasks requiring a different coding between Python 2 and 3, the LLMs would often only consider one of them (e.g., using the API for Python 3 for both the Python 2 and Python 3 cases).

\begin{tcolorbox}[colback=red!5,colframe=red!40!black]
  \textit{Recommendation:} Re-prompting or adding additional information (for instance Chain Of Thoughts \cite{Wei22-cot, li2023structured} or examples of test cases \cite{liu2024codemind}) can help the LLM in the code generation process. This is noted by a respondent of our survey: \enquote{\textit{Most often time after providing a header or function definition, the model can correctly create the method or function}}. For instance, in the cases we observed in our sample set, with the mismatch between the desired version of the library by the user and the obtained version in the LLM's code, providing the concrete version of the libraries can help. By doing so in the prompt, the user can constrain the LLM to generate accordingly. For instance, Guilherme et al. \cite{Vitor23} generated Java test cases for a particular Java/JUnit version by explicitly mentioning it in the prompt. 
\end{tcolorbox}

\textbf{More testing, less trusting:} Even when given enough information, LLMs might generate faulty codes. In that case, having access to test cases to properly test the generated code is important. This however can be quite a manual effort. In a regression test setting, using automated test case generation tools such as EvoSuite \cite{Fraser14} can help. Leveraging LLMs themselves was also shown to be a potential way to generate unit tests \cite{humanevalplus, tang2023chatgpt}. Nonetheless, one should be cautious even when test cases are used. For instance, in the case of \textit{Misinterpretation} or \textit{Missing Corner Case}, the error might not be obvious, and potential existing test cases might not catch the error in the LLMs' generated code. For instance, Liu et al. \cite{humanevalplus} showed that existing test cases in HumanEval dataset \cite{chen2021evaluating}, which encompasses simpler tasks compared to CoderEval, would not catch all existing mistakes in LLMs' generated code. Worse, if no tests are available, users either have to trust the LLMs' generated code or manually read and analyze the obtained code to make sure there are no errors. The first option introduces a high risk depending on the context in which the code should be used while the second one adds a tedious and time-consuming task for the developers, proportional to the code complexity. With the previous points mentioned, it is obvious that LLM-generated code should not be too trusted as it can easily be error-prone. If more experienced developers are less likely to fall into this trap, novice users or people unfamiliar with the particular task requested in the prompt can easily trust generated code, especially if it \enquote{looks} correct, as highlighted in one comment from our survey participants: \enquote{\textit{It is a risk to become complacent to LLM code generation, because sometimes code looks reasonable (in this case), but you might be unaware of all the ways it could fail}}. Some previous studies also pointed out that more novice users tend to over-rely on the code \cite{Scoccia23, Kazemitabaar23, MORADIDAKHEL2023111734} which can have serious negative effects.

\begin{tcolorbox}[colback=red!5,colframe=red!40!black]
\textit{Recommendation:} When possible, users should make use of test cases. Both in the prompt to improve the generation process of the LLMs \cite{liu2024codemind}, as well as to check the generated code for potential errors. In any case, users (especially novice users) should be mindful when reusing LLMs generated code by trusting the solution is correct. One could draw an analogy with StackOverflow where several studies \cite{Fischer17, Zhang18, Verdi22} showed that users tend to rely on those forums while the code present here might not be always reliable.
\end{tcolorbox}

\textbf{Knowing how and when to use LLMs:} In some cases, if too many details are needed for the LLMs to generate a proper code, it might be more straightforward to write the function directly. For instance, Vaithilingam et al. \cite{vaithilingam2022expectation} showed that users leveraging Copilot for coding tasks were not necessarily faster than users without Copilot for the reasons we observed in our study: the code generated might be buggy, oftentimes not in a straightforward way (\textit{Misinterpretation}/\textit{Missing Corner Case}), or the users might have a hard time understanding the logic of the generated code. This is best put by one of our survey's respondents: \enquote{\textit{It depends on whether you know what you are expecting the code to look like. If so, this is easy to spot, and probably you're better off manually writing the code. I have tried to prompt and engineer, but I find that it's a waste of my time. Sometimes, you will get code generated that misinterprets the prompt, but maybe you are not aware of what the generated solution looks like so you have to read the generated code intently or figure out the library calls that are generated to understand that it misunderstood the prompt}}. Moreover, straight code generation might not be recommended in all cases, such as we analyzed in our study. An alternative would be to relax the problem to instead do \textit{code completion} as commented by a survey's participant: \enquote{\textit{In most of my usage scenarios, I generate code with LLM based on existing code. It could be different from letting LLMs generate code from scratch.}}. In that case, the user can start coding the functionality and then use the LLM to complete some lines to lower the workload. The additional context would help the model in making more accurate and less error-prone generations. Even in the case of the generated code containing an error, it will likely be easier for a developer to debug as part of the logic was implemented by said developer and the generated part should also follow a similar logic. This, however, requires some context to work with, which might hamper the usefulness of the method. Another way to mitigate full code generation would be to use LLMs to kick-start coding, saving time on more redundant or simple parts. This has the advantage of not necessarily needing code context and letting the users have more control over the code. This was another point mentioned in the study of Vaithilingam et al. \cite{vaithilingam2022expectation}: even when they did not observe a significant time advantage while using LLMs, developers still preferred them because they could kick-start the coding process and avoid searching for solutions on the Internet.

\begin{tcolorbox}[colback=red!5,colframe=red!40!black]
\textit{Recommendation:} On one hand, if the user has a precise idea of what needs to be done, the advantage of generating the code brought by the LLMs might not offset the time it takes to correctly prompt the LLMs and to debug/understand the generated code by the user. In that case, manual coding can be more recommended. Alternatively, code completion can be used, as the additional code context provided by the users can both guide the LLMs and lead to a code that is more understandable by the users. On the other hand, if the user has no idea how to tackle the task, LLMs can serve as a code kickstart.
\end{tcolorbox}

\subsection{Lessons Learned for Researchers}\label{sec:research}

Following our results and analysis, we provide additional lessons learned that could be future potential research venues for the community regarding bugs in the code generated by LLMs:

\textbf{Repairing the code generated by LLMs:} One way to mitigate bugs in LLM-generated code would be to directly repair them, ideally before the code is returned to the user. Automatic Program Repair (APR) tools \cite{Monperrus18} have been used to fix traditional software programs and several studies \cite{CCTest23, moon2023coffee} started applying them to 
code generated by LLMs. These studies, however, made use of simpler datasets (e.g., HumanEval \cite{chen2021codex} or LeetCode \cite{LeetCode}, which are not based on programming tasks extracted from real-world projects) 
and still failed to find several mistakes. 
Hence, we recommend analyzing existing bugs in practical tasks similarly to what we did, which could be beneficial to guide such repair. This could help improve existing APR tools for LLMs generated code or complement prompt for LLMs to self-check generated code against identified bug patterns.

\textbf{Proposing Code features related Benchmarks for Testing code LLMs:} We observed that LLMs-generated code can contain different bug patterns and that, depending on the code dependency level, LLMs might be more prone to certain errors. Thus, creating standardized benchmark datasets and evaluation metrics specifically tailored for assessing code-related features, the performance of bug detection and triaging methods for LLM-generated code. This will enable fair comparison and evaluation of different testing approaches. Existing benchmarks already focusing on code repair such as CodeXGlue \cite{lu2021codexglue} or HumanEvalPack \cite{Muennighoff2023octopack}, on dependency level such as CoderEval \cite{yu2023codereval} or ClassEval \cite{du2023classeval} or even on code efficiency such as EffiBench \cite{huang2024effibench} are important to be considered. Hence, promoting new benchmarks or complementing the above benchmark with additional features such as different bug patterns, dependency levels or other code-related features could help in better assessing LLMs shortcomings.

\textbf{Interpretable AI for Code Generation:} Besides potential APR techniques, fixing and diagnosing the bugs in LLMs generated code remain manual tasks. This task can be quite intensive depending on the nature of the bug patterns introduced such as \textit{Missing Corner Case} or \textit{Misinterpretation} as we described in our study. As such, employing interpretability techniques for LLMs could help in providing insights into potential fixes or assisting in bug detection. Such techniques analyze how input (i.e., prompt) variations affect a model's output \cite{li2024machines}. By monitoring the effect of presence or absence of different combinations of input elements, these approaches facilitate uncovering the relative contribution of each element to the output, enabling model analysis without necessitating access to its internal architecture \cite{liu2023reliability}. Thus, applying interpretable AI techniques to LLMs to enhance transparency in code generation is also an important research direction. For instance, Ji et al. \cite{Ji2023benchmarking} analyzed how prompt changes impact code generation.

\section{Related Works}\label{sec:related_work}

In this section, we review related works and highlight their findings relevant to buggy code generated by LLMs. While various studies have incorporated LLMs for diverse programming tasks, our focus is specifically on studies that investigated bugs in LLM-generated code. We also discuss studies proposing LLM-based programming assistant tools and studies that examine the quality of code generated by LLMs.

Vaithilingam et al.~\cite{vaithilingam2022expectation} conducted a human study involving 24 participants to investigate the user experience of using Copilot for completing programming tasks. Their study revealed that participants using Copilot had a lower success rate in accurately completing these tasks compared to those using  Intellisense\footnote{\url{https://visualstudio.microsoft.com/services/intellicode/}}. This occurred primarily because the participants struggled to detect and correct 
the errors contained in the code generated by Copilot. Participants faced difficulties correcting the bugs generated by Copilot, to the point that they preferred writing the code from scratch instead of repairing the bugs contained in Copilot's code.

Moradi et al.~\cite{MORADIDAKHEL2023111734} assessed the quality of code generated by Copilot for solving different fundamental algorithmic problems such as searching and sorting. Moradi et al. also compared the quality of code generated by Copilot as an AI pair programming assistant for certain programming tasks with the quality of human-written code. They also examined the effort required to 
repair bugs in code generated by Copilot using an APR tool, comparing it with effort required to fix bugs in human-written code. 
Their results highlight some differences in the cost of repairing buggy Copilot-generated code compared to those generated by humans. Their results also suggest that this difference could be due to Copilot occasionally overlooking specific details in the prompt.

Mastropaolo et al.~\cite{mastropaolo2023robustness} studied the robustness of LLM-based programming assistant tools and investigated the extent to which the input provided to Copilot as a prompt affects the generated output. To conduct this study, they employed a Deep Learning based paraphrasing technique to rephrase task descriptions and assessed the correctness of the rephrased descriptions through manual evaluation by the authors. Their results indicate that rephrasing the task description influences the distribution of solutions generated by Copilot across categories of \textit{failing test}, \textit{passing test}, \textit{syntax error}, and \textit{no solution}. Thus, there is a potential loss in accuracy (moving generated code snippets from \textit{passing test} category to \textit{failing test} or \textit{syntax error}) associated with the description of the prompt.

While the findings of these studies shed light on the bug-proneness of code generated by LLMs, none of them have thoroughly investigated the bug patterns and--or the characteristics of such bugs. Only a handful of studies have delved into the types of bugs observed in the code generated by LLMs, and we will discuss them in the following.

Honarvar et al.~\cite{honarvar2023turbulence} conducted a study to assess the robustness of four different LLMs in addressing what they termed as instances or neighborhood questions. To create a set of neighborhoods for a template question, they replaced variables in 60 different programming tasks with input tests collected from their test oracle and generated different instances for a single programming question. In the evaluation process, rather than just labeling a code as buggy, they assigned a correctness score to indicate the number of instances of a template question successfully handled by the code generated by an LLM. Their findings revealed that there are buggy code snippets generated by LLMs that failed only in a few instances. Subsequently, they categorized the reasons behind failures based on observed error types into classes such as \textit{syntax error}, \textit{runtime error}, \textit{assertion error}, and \textit{wrong number of arguments}.

Fan et al.~\cite{fan2023automated} aimed to enhance the reliability of code generated by Codex by fixing its generated buggy code. They employed both an APR tool and also leveraged Codex to repair the buggy code. To assess the feasibility of applying APR to repair buggy code from Codex, they initially analyzed common mistakes found in solutions generated by Codex. Two authors manually attempted to fix the bugs in the Codex-generated code, creating fixing patches by referencing other human-provided solutions for the same task on Leetcode~\cite{LeetCode}. The authors categorized buggy snippets using a predefined set of defect categories, derived from a benchmark study on human buggy code for programming competition~\cite{tan2017codeflaws}. This classification was inspired by the type of fix required for the buggy code, such as \textit{operator mutation} or \textit{variable mutation}.  They concluded that the buggy code generated by Codex shared similar mutant operators for fixing as those found in human buggy code. However, their results on repairing buggy code revealed that Codex outperformed APR tools (which are also inspired by human bugs) in fixing its own buggy code, and they recommended enhancements to APR tools to overcome this limitation. Their conclusions are not confirmed by our study. Indeed, while \textit{Misinterpretation} and \textit{Missing Corner Case} bug patterns, which relate more to something human developers would do (as validated by the survey in Section \ref{sec:validation}), are still present, multiple non-human like mistakes occur even on a more advanced model like Codex (e.g., \textit{Wrong Attribute}, \textit{Hallucinated Object}). This can be explained by the fact that CoderEval is based on actual programs mined from GitHub and not coding competition snippets such as those extracted from LeetCode contests.

Liu et al.~\cite{liu23} conducted an analysis of code quality issues in the code generated by ChatGPT for a specific set of LeetCode problems~\cite{LeetCode}. They utilized static analysis tools, such as Pylint for Python, to examine common characteristics of the generated code. They applied open card sorting to discuss and group the ChatGPT-generated code based on their results on test cases and the outcome of static analysis tools. 
After grouping the code snippets they assigned a name to each group based on common error patterns, leading to four different categories of quality issues in ChatGPT-generated code: \textit{compilation and runtime errors}, \textit{wrong output} (linked to assertion errors), \textit{code style and maintainability}, and \textit{performance and efficiency}. Within each category, they counted error types from the messages raised by static analysis tools and reported the distribution of error types. Their results revealed that code snippets passing all test cases exhibited fewer quality issues, although they were not entirely free of such issues. While we did not study quality issues (as we focused on bugs rather than code quality issues), the prevalence of \textit{Misinterpretation} and \textit{Missing Corner Case} categories in our dataset are in line with the prevalence of \textit{Wrong Output} category in their study. Thus, we reached similar observations in more practical tasks (compared to LeetCode competitions which are simpler). 
Moreover, their \textit{Compilation and Runtime Error} in the LeetCode tasks match one or more of our categories, for instance, the \textit{Wrong Attribute} or \textit{Wrong Input Type}. 
Many bug categories found in our study were not found in their study, for example, 
\textit{Hallucinated Object} or \textit{Non-Prompted Consideration}. This difference is likely 
due to the nature of the studied tasks: CoderEval tasks are more practical (as they were directly extracted from GitHub) than the programming competition tasks on the LeetCode. This gives way to more complex errors such as \textit{Prompt-biased Code} or \textit{Hallucinated Object} as the docstring is less detailed and the task more complex in CoderEval. 
We believe that our study and presented bug patterns not only confirm their findings but also improve their results by exploring bug patterns in more practical tasks and not limited to the error types reported by static analysis tools.

Jesse et al. \cite{Jesse2023} examined the proneness of LLMs in generating Simple, Stupid Bugs (SStuBs). Stupid bugs are bugs that require single statement fixes but are challenging to locate. To conduct the study, they used a dataset of SStuBs in Java. They used the code before the error line as a prompt and presented it to the LLM and examined whether the model completed the code by regenerating the bug or generating the fix. The results revealed that the LLMs used in their study produced more  SStuBs than the correct code. Additionally, SStuBs generated by Codex took significantly longer to be fixed, underscoring the challenge posed by the prevalence of SStuBs and the potential difficulty in distinguishing between natural and incorrect outputs of LLMs. The study also demonstrated that adding comments to the code can, to some extent, assist the model in generating fewer SStuBs. While the focus of this study was on Java, not all equivalent SStuBs in Python were observed in our sample set. However, we categorized bugs with similar characteristics as \textit{Missing Corner Cases}, where a small change or a single statement modification can transform the buggy code into correct code.


The study by Pan et al.~\cite{pan2023understanding} is focused on analyzing the effectiveness of LLMs in code translation and is the only study that as part of their results developed a systematic bug taxonomy of buggy code generated by LLMs when employed for code translation tasks. The characteristics outlined in their taxonomy are particularly relevant to the challenges associated with translating code from one language to another such as \textit{Mismatch of API behavior between source and target}, \textit{Removal of logic from the source code}, and \textit{Mismatch of behavior after replacing API call}.

To the best of our knowledge, our study is the first to systematically analyze the characteristics of LLM-generated buggy code and construct a bug taxonomy based on the observed bug patterns. Furthermore, we are the first to evaluate the appropriateness of our proposed taxonomy on LLM-generated buggy code through a survey study targeting users of LLMs. In contrast to prior studies that based their analyses on programming tasks collected from platforms like LeetCode, our analysis consists of the buggy code generated by three different LLMs using real-world programming tasks collected from the CoderEval benchmark dataset. We assert that our findings contribute to a more comprehensive understanding of bugs in LLM-generated code.


\section{Threats to Validity}
\label{sec:validity}
\textit{Construct validity:} Our methodology may pose a potential threat as the process of collecting and labeling buggy samples could influence our results. While our methodology is similar to that of existing works that categorized bugs \cite{Tambon2023silent,DL_faults,nikanjam2021DRL}, we addressed this threat by meticulously describing our approach, to allow for external validations. Another limitation could come from the dataset we used: Python functions and the three examined LLMs. Python is a widely used programming language in various domains, and the examined LLMs have been used in previous studies for code generation tasks~\cite{chen2022codet,zan2022neural,nguyen2022empirical,lemieux2023codamosa,chen2023teaching}. In the absence of a pre-existing taxonomy for categorizing bugs, we recognized the risk of introducing bias into our classification. To mitigate this risk, we followed an open coding procedure. Each rater independently evaluated buggy samples, and conflicts were discussed until a consensus was reached. Additionally, to validate our categories, we surveyed practitioners who employed LLMs for code generation. Finally, removing bugs classified as runnable level above \textit{plib\_runnable} could introduce a loss of information regarding the bug patterns identified. Nonetheless, any bugs in the code generated from the prompt of those categories would not necessarily reflect a weakness of LLM in generating code but just a lack of proper information. As such, we preferred removing them and potentially losing some information rather than adding noises to our study.\\
\textit{Internal Validity:} One source of internal threats to validity is the potential bias in manual inspection and labeling of buggy samples. To alleviate this threat, the first two authors (two senior Ph.D. candidates who have experience in SE research and programming) labeled buggy code samples independently. After a series of meetings with the third author (a senior researcher with 10 years of research experience in SE and AI), a consensus was reached on the labeling process and the criteria for categorizing different bug patterns. Another threat arises from the sampling method used to select buggy snippets from  CoderEval. To address potential sampling bias, we followed similar statistical approaches \cite{Zhang19, morovati2024bug} utilizing 95\% confidence levels and 5\% confidence intervals. Furthermore, we validated through a survey, following existing studies \cite{nikanjam2021DRL, Tambon2023silent, DL_faults}, to verify that our obtained categories are representative of bug patterns faced by practitioners. The response rate, 8.9\%, was in line with recent survey studies in SE that received response rates of 8.4\% to 13.3\%  \cite{Nardone23,nikanjam2021DRL,cater2005addressing}. In particular, we showed in Section \ref{sec:validation} that there is a correlation between the bug patterns encountered by respondents and the distribution of bug patterns sampled for Codex, which further comforts us that our sampling is representative. Finally, no new bug patterns were reported by survey respondents.\\
\textit{External Validity:} 
One external threat to the validity is the LLMs used to generate the code in the dataset. In this study, three LLMs were used, CodeGen, PanGu-Coder, and Codex: the first two are open-source and have been used in previous studies that harness the power of LLMs for code generation tasks~\cite{chen2022codet,chen2023teaching,zan2022neural,bogatinovski2023auto}, and the third one, Codex, is the model behind Copilot. Another threat is that the tasks for which the LLMs generated code may not be representative of real tasks which could limit the relevance of the bug patterns found. The choice of the CoderEval dataset mitigates this issue, as it is based on real tasks mined from GitHub. Nonetheless, future works should consider expanding our study to cover a more diverse set of LLMs and functions (problem and prompt) to generate code. Another threat to external validity stems from our focus on \textit{Python} code. Since \textit{Python} is one of the most popular languages for development, we believe that our findings are relevant to the majority of programming languages. However, some language-specific bugs can not be detected in Python because of its properties. For instance, type-based variable declaration or memory management issues would not be something we could easily obtain in a Python program. Thus, future works should consider expanding our study to include other programming languages. The last threat is regarding the rapid development of LLMs and the long-term relevance of the proposed taxonomy. While future studies could expand this taxonomy based on the evolving landscape of LLM advancements, we aimed to draw attention to the distinctive nature of bugs in code generated by LLMs compared to human bugs, even in common types like syntax errors. Moreover, we believe the bug patterns remain broad and will not be overshadowed by the advancements in LLMs over time. Categories such as \textit{Prompt-biased} code,\textit{ Hallucinated Objects}, or \textit{Wrong Input Type} are linked to the limitations arising from the generative nature of LLMs. While they may see a reduction in frequency with improved model performance, they are likely to persist. \\
\textit{Conclusion Validity:} The conclusions drawn in this study have certain limitations, notably the potential for wrong/missing bug patterns. We conducted manual inspections of 333 buggy snippets from a pool of over 1,997 in the dataset. Similarly, buggy snippets were generated by 3 different LLMs to ensure the reliability of our findings. In both instances, the sample sizes were deemed substantial enough to be representative and avoid leading to misguided conclusions. Additionally, to minimize potential errors, bug labeling was independently conducted by two raters and subsequently discussed. Moreover, a validation survey was used to compare our results to users' experience and feedback. Lastly, we have provided a replication package \cite{rep-package} to facilitate the reproducibility of our findings, enabling other researchers to expand upon our study.

\section{Conclusion and Future Work}
\label{sec:conclusion}
In this paper, we empirically investigated the bug patterns observed in code generated by LLMs. Our analysis provides insights into the nature, frequency, and distribution of these bug patterns. We organized our examination of the patterns in a taxonomy for buggy code generated by LLMs. These analyses lead to interesting observations on different bug patterns found in LLM-generated code, such as \textit{Prompt-biased Code}, \textit{Wrong Attribute} and \textit{Hallucinated Object}, which are not common in human buggy code. The cause of several bug patterns is associated with the natural language description of the prompts. Our results indicate that the absence of certain details or even the inclusion of an extra newline in the prompt—typically not problematic to humans, can cause ambiguity for LLMs. Consequently, this ambiguity may lead to the generation of buggy code by LLMs, such as \textit{Missing Corner Case} or \textit{Prompt-biased Code} patterns.

In addition, certain bug patterns originate from the stochastic generative nature of LLMs such as \textit{Hallucinated Object} or \textit{Wrong Attribute} which are also not common in human-written buggy code. While the restricted information of the prompt might explain the existence of \textit{Misinterpretation} and \textit{Missing Corner Case} bug patterns, there is still a non-trivial number of bug patterns (64\% of our sample set) such as \textit{Wrong Attribute} or \textit{Hallucinated Object}, which should not be linked to a lack of information in the prompt. For instance, in Listing \ref{snippet6}, LLM should not apply the `min' function on a list of classes, irrespective of whether or not the docstring describes the expected function enough, as it is simply not applicable to this situation.

We evaluated the relevance of our findings by conducting an online survey targeting software practitioners and researchers involved in LLMs. The respondents of the validation survey mostly encountered the bug patterns we observed. They assessed that, while the majority of the bug patterns can be relatively easy to diagnose and fix, several categories such as \textit{Misinterpretation} or \textit{Non-Prompted Consideration} are difficult to diagnose and fix. Moreover, the patterns of bugs in LLMs generated code are different from the bug patterns of human developers even in categories such as \textit{Syntax Error}.

Our results serve as an initial taxonomy of bug patterns observed in LLM-generated code. While further investigation is required to compare their appearance in human-written code and other LLMs, the findings in this study combined with the validation survey represent the first steps in characterizing bugs in LLM-generated code. This opens up avenues for future studies on enhancing the quality of code generated by LLM-based programming assistant tools. Our findings also can be leveraged to improve different SE tools that rely on characteristics of human buggy code, such as mutant operators in MT or repair rules in APR tools, as well as potential development in Computer Science (CS) education studies.\\

\section*{Acknowledgment}
We would like to thank the participants who contributed to the survey questionnaire.

\section*{Data Availability Statement}
The dataset generated during the current study is available in the replication package, which is accessible via \cite{rep-package}.

\bibliographystyle{IEEEtran}
\bibliography{bibliography}

\section*{Appendix}

\begin{appendix}

\section{Reddit Post}\label{secA1}
\begin{figure}[h]
    \centering
    \includegraphics[width=0.75\textwidth]{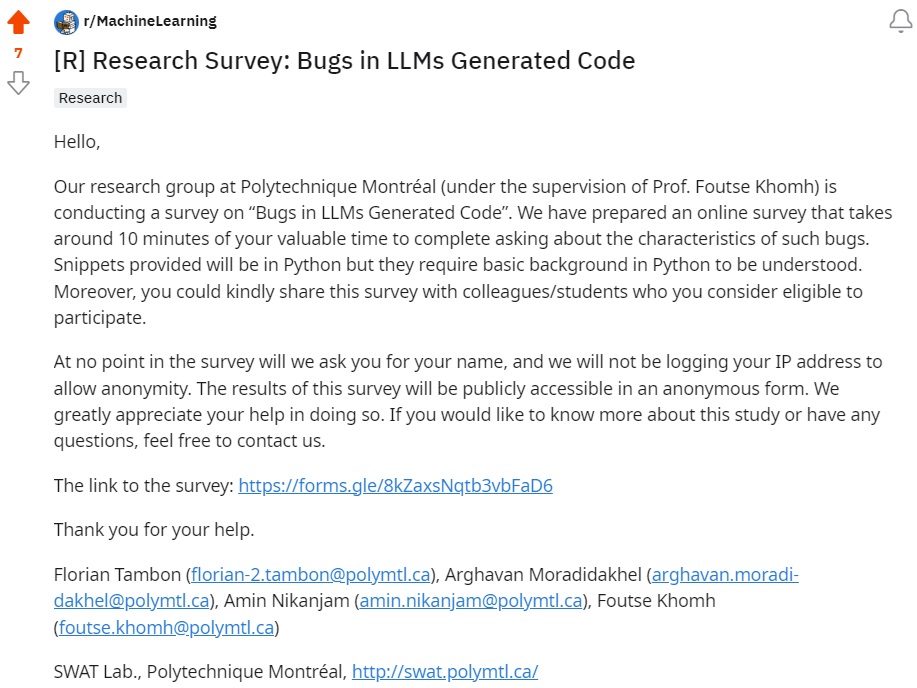}
    \caption{The posted survey on Reddit.}
    \label{fig:reddit_post}
\end{figure}

We posted our survey questionnaire on two relevant Reddit channels: \textit{LocalLLaMA} and \textit{MachineLearning}. Figure~\ref{fig:reddit_post} shows the posted survey on Reddit.

\section{Survey}\label{secA2}
\subsection{The details of the questionnaire in the survey}
\begin{figure}
    \centering
    \includegraphics[width=0.8\textwidth]{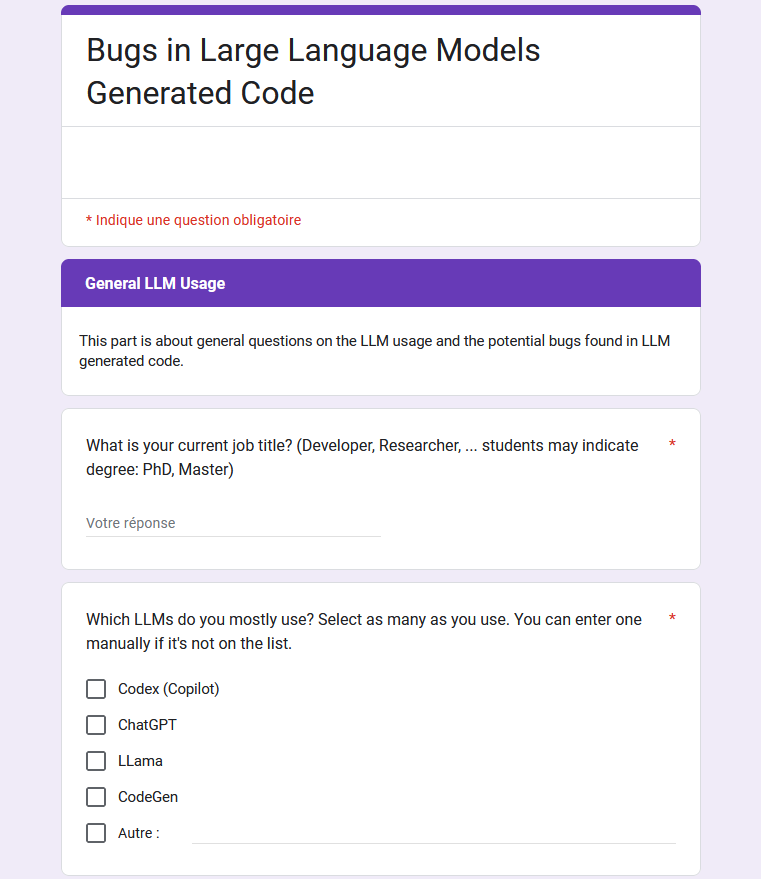}
    \caption{Survey received by the participant}
    \label{fig:survey}
\end{figure}

To prepare the survey, we used Google Forms~\cite{googleForm}, a well-known online tool for creating and sharing surveys and quizzes. The survey has two parts inspired by previous empirical studies on bug patterns~\cite{nikanjam2021DRL,DL_faults,Tambon2023silent}. In the first part, we asked demographic questions to participants i.e. what is your current job title, what LLMs they have used, and for what programming languages they have used LLMs to generate code, other than Python (since it is the programming language of the given code snippets), as presented in Figure~\ref{fig:survey}. Then, we asked for more specific questions about bugs in LLMs generated code, more precisely, we asked:

\begin{itemize}
    \item How often do you encounter mistakes in LLMs generated code you use? Those mistakes can be anything, from code that will not compile (variable not defined, function that does not exist) to more silly choices from the LLMs (multiples If condition checking the same thing or casting back and forth a variable for no reason).
    \item When you encounter mistakes in LLMs generated code, how complex is fixing those issues?
    \item To fix those issues, how do you proceed?
\end{itemize}

 For the first three questions, we used a 5-level Likert scale \cite{oppenheim2000questionnaire} (Never to Always or Trivial to Very Hard). For, the third one, the respondents could choose between \enquote{Manually}, \enquote{Re-prompting the LLM}, \enquote{Combination of both} or \enquote{Other} (and they were given the possibility to write down how).

\begin{figure}
    \centering
    \includegraphics[width=0.8\textwidth]{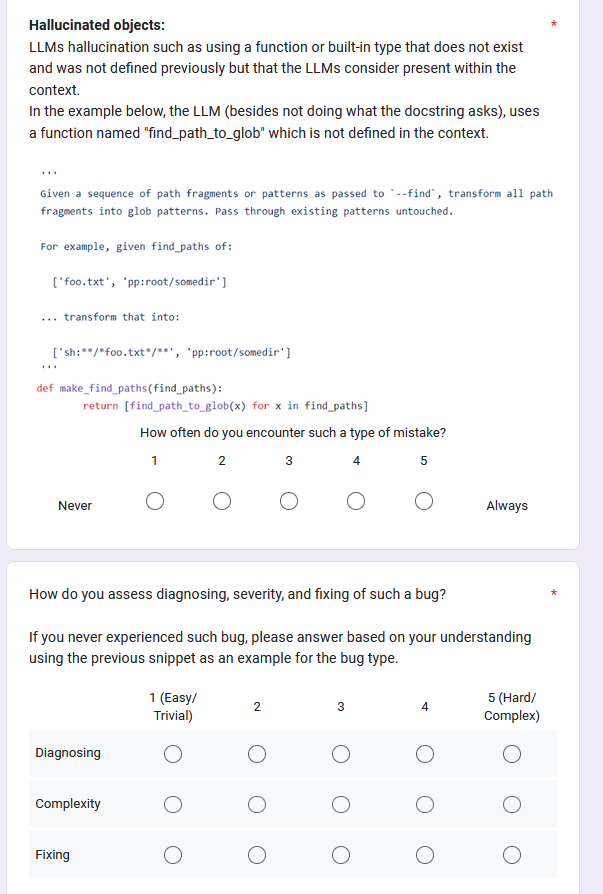}
    \caption{Example of questions for a bug pattern in the survey
    . The description of the task is collected from the CoderEval dataset~\cite{yu2023codereval}.
    }
    \label{fig:survey_example}
\end{figure}

In the second part of the survey, we wanted to consolidate our bug patterns and gather the participants’ feedback on several aspects of each pattern of bug in our study. An example of questions in the second part is shown in Figure~\ref{fig:survey_example}. The structure of the questions is similar for each pattern of bugs in the taxonomy.

\subsection{Survey Results}\label{secA3}

\begin{sidewaystable}
\centering
\caption{Results of validating survey for the bug patterns. Questions related to the frequency of encounter of bug patterns, the difficulty to diagnose and fix them as well as the complexity of the bug. We highlight in bold the highest number in each category for each bug pattern. 1 represents never/easy/trivial/low and 5 represents always/hard/complex/high.
}
\resizebox{\textwidth}{!}{
\begin{tabular}{|c|c|c|c|c|c|}
\hline
\multirow{3}{*}{\textbf{Types of Bug}} & \multicolumn{4}{c|}{\textbf{Responses}}\\
& Frequency & Diagnosing & Complexity & Fixing \\
& 1 | 2 | 3 | 4 | 5 & 1 | 2 | 3 | 4 | 5 & 1 | 2 | 3 | 4 | 5 & 1 | 2 | 3 | 4 | 5\\

\hline
\hline
Hallucinated & \multirow{2}{*}{$17.6\%$ - $\textbf{26.5}\%$ - $23.5\%$ - $23.5\%$ - $8.8\%$} & \multirow{2}{*}{$\textbf{79.4}\%$ - $17.6\%$ - $0\%$ - $0\%$ - $2.94\%$} & \multirow{2}{*}{$\textbf{47.1}\%$ - $32.4\%$ - $14.7\%$ - $5.9\%$ - $0\%$} & \multirow{2}{*}{$17.6\%$ - $26.5\%$ - $\textbf{29.4}\%$ - $20.6\%$ - $5.9\%$} \\
object & & & &\\
\hline
Incomplete & \multirow{2}{*}{$20.6\%$ - $11.8\%$ - $29.4\%$ - $\textbf{32.4}\%$ - $5.9\%$} & \multirow{2}{*}{$\textbf{73.5}\%$ - $14.7\%$ - $2.9\%$ - $8.8\%$ - $0\%$} & \multirow{2}{*}{$\textbf{41.2}\%$ - $38.2\%$ - $8.8\%$ - $11.8\%$ - $0\%$} & \multirow{2}{*}{$26.5\%$ - $20.6\%$ - $11.8\%$ - $\textbf{29.4}\%$ - $11.8\%$} \\
Generation & & & & \\
\hline
\multirow{2}{*}{Misinterpretation} & \multirow{2}{*}{$5.9\%$ - $\textbf{35.3}\%$ - $26.5\%$ - $26.5\%$ - $5.9\%$} & \multirow{2}{*}{$29.4\%$ - $17.6\%$ - $\textbf{32.4}\%$ - $17.6\%$ - $2.9\%$} & \multirow{2}{*}{$14.7\%$ - $23.5\%$ - $26.5\%$ - $\textbf{32.4}\%$ - $2.9\%$} & \multirow{2}{*}{$5.9\%$ - $11.8\%$ - $\textbf{38.2}\%$ - $35.3\%$ - $8.8\%$} \\
& & & & \\
\hline
\multirow{2}{*}{NPC} & \multirow{2}{*}{$17.6\%$ - $23.5\%$ - $\textbf{35.3}\%$ - $20.6\%$ - $2.9\%$} & \multirow{2}{*}{$23.5\%$ - $23.5\%$ - $\textbf{26.5}\%$ - $23.5\%$ - $2.9\%$} & \multirow{2}{*}{$17.6\%$ - $\textbf{32.4}\%$ - $29.4\%$ - $14.7\%$ - $5.9\%$} & \multirow{2}{*}{$\textbf{29.4}\%$ - $\textbf{29.4}\%$ - $23.5\%$ - $11.8\%$ - $5.9\%$} \\
& & & & \\
\hline
Silly & \multirow{2}{*}{$\textbf{29.4}\%$ - $\textbf{29.4}\%$ - $20.6\%$ - $17.6\%$ - $2.9\%$} & \multirow{2}{*}{$\textbf{58.8}\%$ - $26.5\%$ - $14.7\%$ - $0\%$ - $0\%$} & \multirow{2}{*}{$\textbf{67.6}\%$ - $29.4\%$ - $2.9\%$ - $0\%$ - $0\%$} & \multirow{2}{*}{$\textbf{55.9}\%$ - $26.5\%$ - $8.8\%$ - $8.8\%$ - $0\%$} \\
Mistake & & & & \\
\hline
Wrong & \multirow{2}{*}{$17.6\%$ - $\textbf{38.2}\%$ - $14.7\%$ - $20.6\%$ - $8.8\%$} & \multirow{2}{*}{$\textbf{35.3}\%$ - $23.5\%$ - $20.6\%$ - $17.6\%$ - $2.9\%$} & \multirow{2}{*}{$17.6\%$ - $\textbf{50}\%$ - $14.7\%$ - $17.6\%$ - $0\%$} & \multirow{2}{*}{$26.5\%$ - $\textbf{38.2}\%$ - $17.6\%$ - $14.7\%$ - $2.9\%$} \\
Input Type& & & &\\

\hline
Missing & \multirow{2}{*}{$11.8\%$ - $17.6\%$ - $20.6\%$ - $\textbf{35.3}\%$ - $14.7\%$} & \multirow{2}{*}{$5.9\%$ - $14.7\%$ - $26.5\%$ - $\textbf{44.1}\%$ - $8.8\%$} & \multirow{2}{*}{$14.7\%$ - $14.7\%$ - $\textbf{38.2}\%$ - $23.5\%$ - $8.8\%$} & \multirow{2}{*}{$17.6\%$ - $\textbf{29.4}\%$ - $17.6\%$ - $\textbf{29.4}\%$ - $5.9\%$} \\
Corner Cases & & & &\\
\hline

Wrong & \multirow{2}{*}{$0\%$ - $17.6\%$ - $26.5\%$ - $\textbf{47.1}\%$ - $8.8\%$} & \multirow{2}{*}{$20.6\%$ - $\textbf{44.1}\%$ - $26.5\%$ - $5.9\%$ - $2.9\%$} & \multirow{2}{*}{$17.6\%$ - $\textbf{50}\%$ - $17.6\%$ - $11.8\%$ - $2.9\%$} & \multirow{2}{*}{$8.8\%$ - $\textbf{38.2}\%$ - $35.3\%$ - $11.8\%$ - $5.9\%$} \\
Attribute & & & &\\
\hline

Syntax & \multirow{2}{*}{$\textbf{47.1}\%$ - $20.6\%$ - $20.6\%$ - $5.9\%$ - $5.9\%$} & \multirow{2}{*}{$\textbf{85.3}\%$ - $8.8\%$ - $5.9\%$ - $0\%$ - $0\%$} & \multirow{2}{*}{$\textbf{82.4}\%$ - $8.8\%$ - $8.8\%$ - $0\%$ - $0\%$} & \multirow{2}{*}{$\textbf{70.6}\%$ - $14.7\%$ - $2.9\%$ - $11.8\%$ - $0\%$} \\
Error & & & &\\
\hline

Prompt-biased & \multirow{2}{*}{$11.8\%$ - $23.5\%$ - $11.8\%$ - $\textbf{44.1}\%$ - $8.8\%$} & \multirow{2}{*}{$\textbf{50.0}\%$ - $26.5\%$ - $17.6\%$ - $5.9\%$ - $0\%$} & \multirow{2}{*}{$11.8\%$ - $\textbf{47.1}\%$ - $26.5\%$ - $11.8\%$ - $2.9\%$} & \multirow{2}{*}{$5.9\%$ - $23.5\%$ - $26.5\%$ - $\textbf{29.4}\%$ - $14.7\%$} \\
Code & & & &\\
\hline
\end{tabular}}
\label{table:table-survey-taxonomy}
\end{sidewaystable}

\end{appendix}


\end{document}